\documentclass[10pt,journal,a4paper]{IEEEtran}
\usepackage{amssymb,amsmath}
\usepackage{algorithm}
\usepackage{scalerel}
\usepackage{dsfont}
\usepackage[table]{xcolor}
\usepackage{gensymb}
\usepackage{mathbbol}
\usepackage{bbm}
\usepackage{algpseudocode}
\usepackage{cite}
\usepackage{graphicx,subfigure}
\usepackage{multirow}
\usepackage{psfrag}
\usepackage{url}
\usepackage{xcolor}
\usepackage[absolute,overlay]{textpos}
\usepackage{algpseudocode}
\usepackage{float}
\usepackage{lettrine}
\usepackage{indentfirst}
\usepackage{amsthm}

\newtheorem{theorem}{Theorem}

\usepackage{array}
\usepackage{psfrag}
\usepackage{url}
\usepackage{caption}
\usepackage[caption=false]{subfig}
\usepackage[font=footnotesize]{caption}
\usepackage{subfigure}

\makeatletter
\def\thickhline{%
  \noalign{\ifnum0=`}\fi\hrule \@height \thickarrayrulewidth \futurelet
   \reserved@a\@xthickhline}
\def\@xthickhline{\ifx\reserved@a\thickhline
               \vskip\doublerulesep
               \vskip-\thickarrayrulewidth
             \fi
      \ifnum0=`{\fi}}
\makeatother

\newlength{\thickarrayrulewidth}
\begin{document}
	
	\title{Extreme Value Theory-enhanced Radio Maps for Handovers in Ultra-reliable Communications
}
	\author{
        \IEEEauthorblockN{Dian Echevarría Pérez, \IEEEmembership{Graduate Student Member, IEEE},  %
        Onel L. Alcaraz López, \IEEEmembership{Senior Member, IEEE}, Hirley Alves, \IEEEmembership{Member, IEEE}
        } 
        
		\thanks{The authors are with the Centre for Wireless Communications (CWC), University of Oulu, Finland. \{dian.echevarriaperez,  onel.alcarazlopez, hirley.alves\}@oulu.fi}

        \thanks{This work was supported by 6G Flagship (Grant Number 369116) funded by the Research Council of Finland, and the Finnish Foundation for Technology Promotion.}
    }  
    \maketitle

    \renewcommand\thefootnote{}% clear numbering
\footnotetext{%
© 2025 IEEE. Personal use of this material is permitted. Permission from IEEE must be obtained for all other uses, in any current or future media, including reprinting/republishing this material for advertising or promotional purposes, creating new collective works, for resale or redistribution to servers or lists, or reuse of any copyrighted component of this work in other works.}%
\renewcommand\thefootnote{\arabic{footnote}}% restore numbering

	\begin{abstract}
   Efficient handover (HO) strategies are essential for maintaining the stringent performance requirements of ultra-reliable communication (URC) systems. This work introduces a novel HO framework designed from a physical-layer perspective, where the decision-making process focuses on determining the optimal time and location for performing HOs. Leveraging extreme value theory (EVT) and statistical radio maps, the proposed method predicts signal behaviour and enables efficient resource allocation. The framework ensures seamless HOs and improved system performance by facilitating effective resource transitions and coordination across spatial locations while incorporating mechanisms to mitigate the ping-pong effect. Comparative evaluations demonstrate that this strategy provides superior service availability and energy efficiency than traditional HO mechanisms, highlighting its effectiveness in URC environments. 
	\end{abstract}

\section{Introduction}
Ultra-reliable communication (URC) represents a key operation mode in current and future wireless communications networks. Achieving high reliability poses a big challenge, requiring, for instance, an accurate characterisation of the distribution tail of the signal-to-interference-plus-noise ratio (SINR) ~\cite{Mahmood.2020,lopez2022statistical}. Although difficult in practice, a complete characterisation of the SINR tail enables efficient resource allocation methods that guarantee reliability requirements, leading to fewer errors/retransmissions, thus also reducing transmission latency. Such characterisation can be significantly enhanced if the spatial structure/correlation of the environment is exploited. Indeed, the so-called channel/radio maps offer a detailed spatial representation of signal parameters across different geographical areas, further improved when combined with extreme value theory (EVT) for extreme/rare event characterisation \cite{perez2024evt}. 

Meanwhile, developing effective handover (HO) strategies between base stations (BSs) is essential to ensure consistent quality-of-service (QoS) under system dynamics triggered, \textit{e.g.,} by the movement of the user equipments (UEs) and/or other scene updates. Leveraging the spatial characteristics of the environment and the device's position can significantly improve these strategies. This is particularly important for URC applications, where constant monitoring of quality metrics may not be practical, as it adds signaling overhead, increases delays, consumes additional energy, and can result in excessive handovers, leading to inefficiencies.
 
\subsection{EVT and radio maps for URC}

EVT and radio maps have been widely exploited for resource allocation in URC systems. These tools have been applied either separately or in combination to address the challenges of guaranteeing reliability under limited or spatially varying channel knowledge. EVT was employed in \cite{perez2023extreme} to develop a minimum-power precoding scheme for a multi-UE downlink (DL) scenario with strict outage probability constraints. The authors analysed the impact of estimation error sample size on system performance and showed that there exists an optimal pilot sequence length that minimises transmit power. EVT-based channel modelling for URC was explored in \cite{EVTMehrnia}, where the authors proposed a method to characterise the statistics of the multivariate tail distribution of received power in a MIMO system, demonstrating improved tail fitting over extrapolation-based approaches. In \cite{mehrnia2022extreme}, EVT was used to design a rate selection scheme by fitting the lower tail of the received power distribution to the generalised Pareto distribution (GPD), enabling the determination of maximum transmission rates that satisfy reliability constraints. Additionally, EVT and federated learning were combined in \cite{zheng2022data} to model and learn the distribution of extreme queue lengths, leading to a resource allocation mechanism that minimises the worst-case queue length in vehicular networks. On the other hand, radio maps have been leveraged to support URC, as demonstrated in \cite{kallehauge2022predictive}, where Gaussian processes were used to generate quantile maps of the logarithmic signal-to-noise ratio (SNR) at unobserved locations, based on measurements at known spatial positions. These maps were then employed to design a rate selection scheme suitable for ultra-reliable communication. The study in \cite{10105152} evaluated the use of radio maps and channel charting in the context of URC, demonstrating that radio maps outperform channel charting in terms of outage probability and spatial prediction accuracy of channel capacity. More recently, \cite{perez2024evt} proposed a unified framework that integrates EVT and radio maps for resource allocation under URC requirements. In this work, Gaussian processes were used to construct spatial maps of the GPD parameters by leveraging environmental correlation. The proposed approach achieved higher transmission rates and improved service availability compared to methods relying solely on SNR quantile-based radio maps. The authors in \cite{10879538} proposed a framework that combines radio maps with non-parametric models and EVT to estimate rare-event channel statistics within a Bayesian formulation. They applied this approach to a rate selection problem aimed at supporting URC services.

\subsection{Location and machine learning-based HO strategies}\label{section I-B}
Classical HO methods typically rely on monitoring QoS metrics such as received signal strength (RSS) or SINR, triggering an HO when these metrics fall below a predefined threshold for a duration exceeding the time-to-trigger (TTT). Depending on service requirements, additional criteria like throughput, load balancing of BSs, or UE speed may also influence the HO decision process \cite{mollel2020intelligent}. Since the HO procedure involves signaling exchanges between BSs and the UE, several positioning-based and machine-learning approaches have been developed to streamline this process and reduce HO preparation time. For example, \cite{kastell2004performance} proposed a localisation-based mechanism using global system for mobile communication (GSM) to accelerate HO preparation, reducing the time needed to identify the next cell. Similarly, \cite{inzerilli2008location} introduced a localisation-based HO scheme aimed at mitigating the ping-pong effect by initiating HOs based on UE location, which performed better than classical power-based methods in terms of reducing HO frequency. 

In high-speed scenarios, \cite{chen2014location} proposed a location-based HO strategy that dynamically computes UE velocity and distance to the BS, combining this information with RSS to trigger preemptive HOs. An RSS prediction-based multi-connectivity scheme was introduced in \cite{zhao2018received}, using predicted RSS values to select the target BS, achieving up to a 50\% reduction in HO frequency compared to traditional methods. Other machine-learning-based approaches include \cite{guo2020joint}, which presented a multi-agent reinforcement learning algorithm for HO management and power allocation to maximise throughput, and \cite{khosravi2020learning}, which proposed a joint HO and beamforming strategy to maintain consistent transmission rates in millimeter-wave systems. Similarly, \cite{alkhateeb2018machine} introduced a proactive HO mechanism where BSs predict future link blockages with nearly 95\% accuracy, enabling timely HO decisions in URC scenarios.

Despite significant advancements, state-of-the-art methods face challenges such as excessive signaling overhead, increased latency, and/or energy inefficiencies stemming from continuous QoS monitoring. Furthermore, these approaches often lack integration with resource allocation strategies that are specifically tailored to meet the stringent reliability and latency demands of URC. To overcome these limitations, this work introduces a novel HO strategy that utilises pre-constructed radio maps and EVT to ensure URC reliability while optimising resource utilisation, including minimising power consumption. Importantly, our approach determines the optimal time and location for HOs based solely on physical-layer metrics, without incorporating factors like BS load or backhaul capacity. While these factors are excluded to establish a robust and focused mechanism at the physical layer, they can be integrated as needed to extend the framework.
\subsection{Contributions}
To the best of our knowledge, this is the first work to present an HO strategy explicitly designed for resource allocation in the URC regime. Building on the radio map construction methodology from \cite{perez2024evt,kallehauge2022predictive}, we extend its application to dynamic HO decisions, leveraging physical-layer information to avoid excessive QoS monitoring. Specifically, our contributions are four-fold\footnote{This work is based on our invention described in \cite{Dian_patent}. The insights and contributions presented herein extend and elaborate upon the foundational concepts.}: 

\begin{itemize}
    \item We propose a novel HO strategy that integrates pre-constructed radio maps and UE location information to optimise resource allocation in the URC regime. While the radio maps are constructed using EVT and Gaussian processes as in prior work\cite{perez2024evt}, we extend their use to support seamless and energy-efficient HO decisions across shared BS coverage areas.

    \item We propose a combination of filtering and peak detection for HO maps with smoother transitions when allocating resources at adjacent spatial positions. Notably, the filtered maps also contribute to a more efficient BS selection and help to prevent the ping-pong effect between BSs when serving moving UEs. 
    \item We introduce control strategies to optimise power usage and service time, reducing the ping-pong effect. Additionally, maximum transmit power constraints are integrated to enhance decision-making, ensuring reliable connectivity and seamless handovers in dynamic conditions.
    \item We assess the performance of the presented approach for a transmit power minimisation problem with outage constraints and show that the energy consumption is highly reduced when compared with other strategies such as serving the UE from the nearest BS. Also, the presented scheme achieves higher service availability than classical HO strategies.   
\end{itemize}
The work is structured as follows. Section \ref{Sect_System} describes the system model and main assumptions. In Section \ref{section3}, we present the EVT-based constraint reformulation and radio map construction. In Section \ref{Section4}, we further process the obtained radio maps and present the BS selection policy. Section \ref{section5} contains the numerical results and validates the proposed algorithms. Finally, \ref{section6} concludes the paper.

\textbf{Notation:} Superscript $(\cdot)^T$ denotes the transpose operator and $(\cdot)^{-1}$ represents the matrix inverse operation. Moreover, $\mathcal{N}(\mathbf{v},\mathbf{R})$ denotes a Gaussian distribution with mean vector $\mathbf{v}$ and covariance matrix $\mathbf{R}$, and $\mathbb{E}[\cdot]$ is the expectation operator. Also, $\Gamma(\cdot)$ and $B_{\nu}(\cdot)$ depict the Gamma function and modified Bessel function of the second kind and order $\nu$, respectively. Finally, $Q^{-1}(\cdot)$ represents the inverse Q-function, $\mathcal{Q}(c,D)$ depicts the $c\%$-quantile operator of the sample set $D$, $\lceil \cdot\rceil$ represents the ceiling operator, and $*$ denotes convolution.  Table \ref{table_0} summarises the main symbols used throughout the paper.

\begin{table}[t!]
    \centering
    \caption{Main symbols used throughout the paper}
\label{table_0}    \begin{tabular}{l l}
        \hline
        \textbf{Symbol} & \textbf{Definition} \\
        \hline
        $M$ & Number of locations with SINR measurements \\
          $B/B'$ & Number of serving/interfering BSs\\
        $N$ & Number of independent i.i.d. SINR measurements \\
        $\gamma_b$ & Achieved SINR\\  $\upsilon^2$ & Noise variance \\
        $h$ & Complex channel coefficient\\
        $O$ & Outage probability\\
        $\gamma_0$ & SINR target \\
        $\zeta$ & Target outage probability \\
        $\mu$, $\xi$, $\sigma$ & Threshold, shape, and scale parameters of the GPD \\
        $\rho$ & Quantile value of the samples\\
        $p_b/p_b'$ & Transmit power of BS $b/b'$\\
        $p_0$ & SINR measurements\\
        $\kappa_b$ & Power scaling factor\\ 
        $T_i$ & Time instant $i$\\
        $\Delta T$ & Hold timer\\ 
        $\Delta p$ & Hysteresis margin\\
        $\tau$ & Quantile of the threshold of the GPD\\
        $\theta$ & Hyperparamter of the kernel\\
        $C_{\text{x}}$, $C_{\text{y}}$ & Radius of the kernel along the x and y axes, respectively\\
        \hline
    \end{tabular}
\end{table}
\section{System model}\label{Sect_System}
\begin{figure}[t!]
    \centering  \includegraphics[width=\columnwidth]{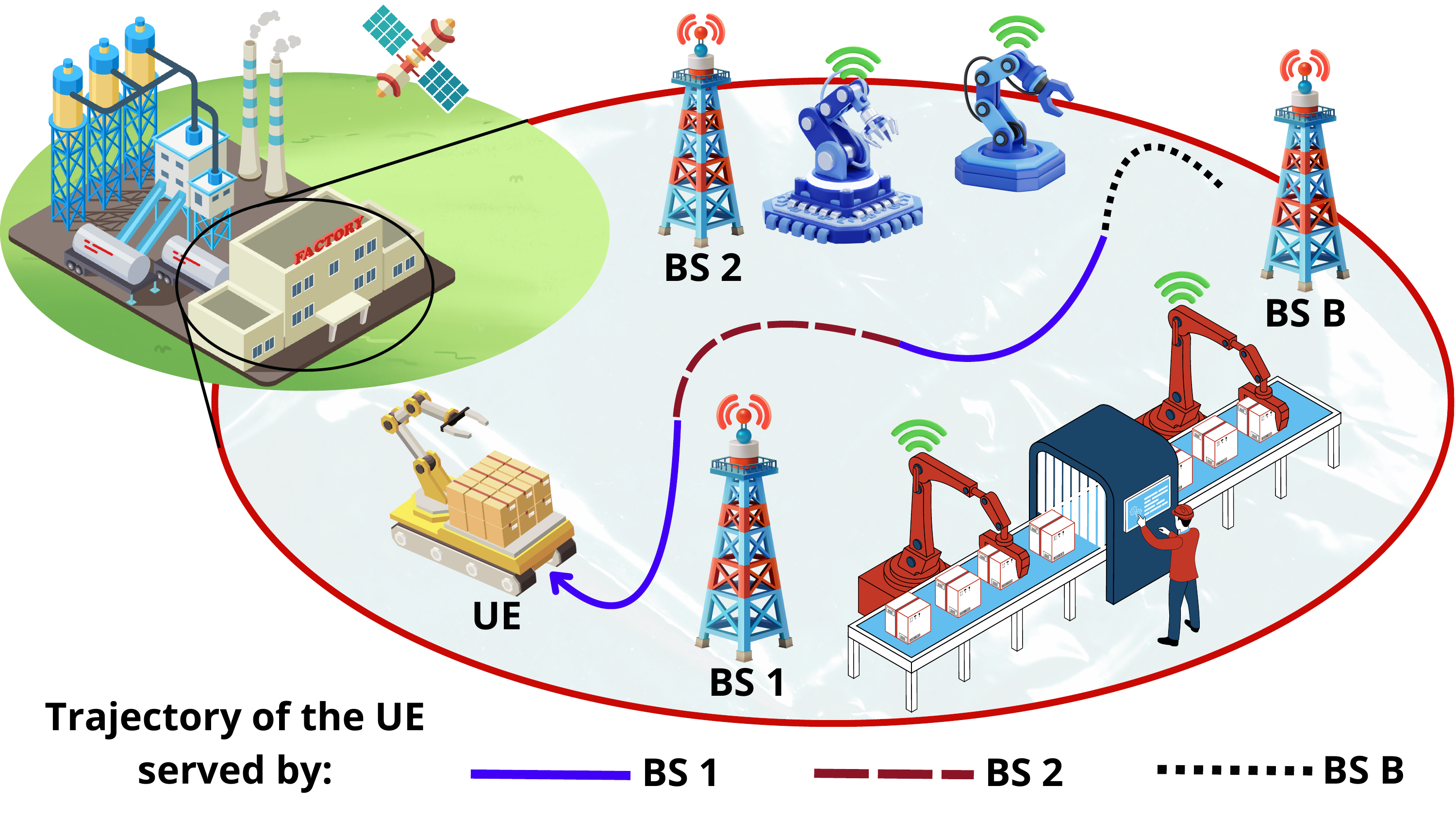}
    \caption{The BSs are serving a moving UE in the DL while guaranteeing URC QoS demands throughout its trajectory in an industrial environment. SINR measurements are collected from past network instances from the sensors deployed within the factory or from UEs at different locations. External interfering BSs are not displayed in the figure.}
    \label{syst_model}
\end{figure}

We consider an indoor industrial environment with $B$ single-antenna small BSs\footnote{Notice that using single-antenna BSs/UEs serves as a baseline. Obviously, exploiting multiple antennas would further improve the performance of the approaches proposed later.} serving single-antenna moving URC UEs in the DL channel within their shared coverage area as shown in Fig. \ref{syst_model}. Each BS $b$ knows $N$ independent and identically distributed (i.i.d.) SINR measurements $\Gamma_b(l_{m}) = \{\gamma_{b,1}(l_{m}), \gamma_{b,2}(l_{m}),...,\gamma_{b,N}(l_{m})\}$ from $M$ different locations $l_{m}= \{x_{m}, y_{m}\}$, $m = 1,2,...,M$. Assume that the BSs perfectly know the $M$ locations and that the channel measurements \textit{i)} do not necessarily coincide for all BSs, \textit{ii)} may be acquired from pre-deployed sensors/devices with a known location, or high-precision systems such as Real-Time Kinematic (RTK) positioning and carrier phase-based positioning \cite{9566601,10015771}, and \textit{iii)} are grouped as  $\mathcal{L}_b=\{l_{1},l_{2},...,l_{m} \}$. Assuming that a UE is served by BS $b$ at a given frequency, time interval, and location $l_{m}'$ in the coverage area with $\mathcal{L}' = \{l_1', l_2', \dots, l_{M'}'\}$ and $M'$ as the number of unobserved locations, the received signal is given by
\begin{equation}\label{received_signal}
y_b(l_{m}') = \sqrt{p_b}h_b(l_{m}')x_b + \sum_{b'=1}^{B'}\sqrt{p_{b'}}h_{b'}(l_{m}')x_{b'}+\textit{v},    
\end{equation}
where $p_b/p_{b'}$ represents the power transmitted by BS $b/b'$, $h_b/h_{b'}(l_m')$ is the complex channel coefficient from BS $b/b'$ to a UE location $l_{m}'$ including small-scale and large-scale effects, $x_b/x_{b'}$ is the transmitted symbol from BS $b/b'$ such that $\mathbb{E}[x_b]=\mathbb{E}[x_{b'}] = 1$. Note that $b' =1,2,\cdots, B'$, where $B'$ denotes the number of external interfering BSs that share the same time-frequency resources as BS $b$.
Finally, 
\textit{v}$\sim\!\!\mathcal{N}(0,\upsilon^2)$ depicts the additive white Gaussian noise with zero mean and variance $\upsilon^2$. Therefore, the perceived SINR is given by
\begin{equation}\label{SNR-eq}
\gamma_b(l_{m}') = \frac{p_b|h_b(l_{m}')|^2}{\sum_{b'=1}^{B'}p_{b'}|h_{b'}(l_{m}')|^2 + \upsilon^2}.    
\end{equation}

\subsection{Problem definition}

As discussed in Section~\ref{section I-B}, an HO typically occurs when a specific QoS metric, such as SINR or RSS, falls below a predefined threshold for a period longer than the TTT. This approach requires continuous monitoring of the quality metric, which can introduce persistent signaling and delays that negatively impact applications with strict latency requirements. To overcome these challenges, alternative HO strategies can be developed to optimise network resource usage by incorporating real-time localisation of the UE. Using location information can simplify the HO decision process and enable more efficient transitions between BSs. For instance, an HO may be initiated to reduce power consumption or improve data rates, even when the current QoS is still acceptable. However, these strategies must include additional mechanisms to prevent excessive switching between BSs within a short timeframe, which could otherwise lead to system instability.

In the following, we focus on designing HO strategies for resource allocation in the URC regime. Specifically, we concentrate our efforts on the transmit power minimisation problem under strict outage constraints. Nevertheless, the presented methodology can be applied to any resource allocation problem. It is important to note that we focus exclusively on the HO strategy at the physical layer, assuming that once a handover is triggered, it is successful. Factors that may affect the success of an HO, such as BS load, backhaul capacity, or interference from neighbouring cells, are not considered in this work. The primary objective of the proposed strategy is to determine the optimal timing and conditions for reducing transmit power, and consequently, energy consumption, while guaranteeing the URC requirements.

The problem can be formulated as 
\begin{subequations}\label{P1}
	\begin{alignat}{2}
	\mathbf{P1:} \ &\underset{\mathbf{p}}{\mathrm{minimise}}  &\ \ \ &\
	\mathbf{p}^T(l_{m}')\mathbf{a}(l_{m}', T_i) \label{P1:a}\\
	&\text{subject to}   &      & \mathcal{O}(l_{m}')\!\triangleq\!\text{Pr}\big\{\gamma_b(l_{m}')\!<\!\gamma_0\big\}\!\le\!\zeta, \forall l_{m}', \label{P1:b}
	\end{alignat}
\end{subequations}
where $\mathbf{p} = [p_1, p_2, ... , p_B]^T$ contains the transmit power of each BS $b$ and $\mathbf{a}$ is a $B\times 1$ predefined activation function with `1' in the position of the active BS serving the UE at location $l_{m}'$ and time instant $T_i$, and `0' in the remaining ones. Also,  $\gamma_0$ depicts the target SINR and $\zeta$ the target outage probability. Notice that the outage probability $\mathcal{O}$ is a widely used metric for evaluating the reliability of wireless communication systems \cite{lopez2022statistical}. The problem $\textbf{P1}$ is nonconvex in its current form, and obtaining a closed-form expression for constraint~\eqref{P1:b} would require exact knowledge of the SINR distribution, which is not available. This requirement is often impractical, as determining the distribution of the SINR would involve fitting the observed data to multiple candidate distributions. Such fitting procedures are typically inaccurate and may not yield the true underlying distribution, thereby limiting the reliability of any resulting expressions. 

\section{Problem reformulation and radio map construction}\label{section3}
\subsection{EVT-reformulation}

EVT is grounded in the principle that the statistical behavior of extreme values in a dataset can be approximated by a limiting distribution, independent of the original distribution from which the data is drawn. In the case of SINR, outage events are associated with rare and exceptionally low values. EVT allows us to characterise this lower tail behavior reliably, even when only a limited number of SINR samples are available. The full distribution is unknown. Thus, as in \cite{lopez2022statistical,perez2023extreme,perez2024evt}, we proceed to reformulate the outage expression in \eqref{P1:b}, exploiting the following EVT theorem.
\begin{theorem}
\label{th:exceedance-th-coles}
(Theorem of Exceedances over Thresholds \cite{coles2001introduction}) For a random variable (RV) $X$ from a non-degenerative distribution and for a large enough threshold $\mu$, the cumulative distribution function (CDF) of $Z =X-\mu$ conditioned on $X>\mu$ converges to
\begin{align}\label{GPD}
    F_Z(z) = 1-\bigg[1+\frac{\xi z}{\sigma}\bigg]^{-\frac{1}{\xi}},
\end{align}
defined on $\{z: z>0 \ \text{and} \ 1+\xi z/\sigma > 0\}$. The distribution in \eqref{GPD} is known as the GPD with shape and scale parameters $\xi$ and  $\sigma$, respectively.
\end{theorem}

 While the parameters $\sigma$ and $\xi$ are typically estimated via log-likelihood methods, the threshold parameter $\mu$ may be determined using the \textit{mean residual life} method or through a complementary approach known as the \textit{parameter stability method} \cite{coles2001introduction}. Another common strategy is the \textit{fixed threshold approach}, wherein the threshold is specified before model fitting. A well-known example of this is the rule proposed by DuMouchel, in which the threshold is chosen as up to the upper 10\% of the data. In this case, $\mu$ can be estimated directly using quantile-based methods, \textit{i.e.}, $\hat{\mu} = \mathcal{Q}(\rho \times 100, X)$ for $\rho \ge 0.9$ \cite{dumouchel1983estimating}. Multiple observations of the SINR at $M$ locations allow us to exploit the theorem to come up with $M$ outage probability expressions that characterise the tail of the distribution of the SINR at each location. 
Notice that SINR samples with low values, corresponding to more severe propagation conditions, lie in the left tail of the distribution. Some observations may also be significantly separated from the main body of the data, which can distort the GPD fit. To address this, we apply a convex transformation \(f(\cdot)\) to each set \(\Gamma_b(l_m)\). This transformation maps the data from the left tail to the right tail and preserves its ordering while compressing extreme values, resulting in more stable threshold selection and improved reliability of the GPD parameter estimates \cite{lopez2022statistical}. Then, we can compute the excess data as
 
\begin{equation}\label{excess}
\psi(l_{m}) = f\big(\Gamma_b(l_{m})\big)-\hat{\mu}(l_{m})\ \big|\ f\big(\Gamma_b(l_{m})\big)>\hat{\mu}(l_{m}),
\end{equation}
where the threshold is determined by applying DuMouchel's rule as
\begin{equation}\label{threshold_eq}
    \hat{\mu}= \mathcal{Q}\Big(\rho\times 100,f\big(\Gamma_b(l_{m})\big)\Big).
\end{equation}
 Moreover, performing log-likelihood estimation\footnote{Notice that 
 maximum likelihood estimation typically involves numerical optimisation to compute parameter estimates \cite{coles2001introduction}.}
 over $\psi(l_{m})$ for finding estimates $\hat{\sigma}$ and $\hat{\xi}$, the outage expression in \eqref{P1:b} can be re-written as \cite{perez2024evt,perez2023extreme}
\begin{equation}\label{Outage_eq}
    \mathcal{O}(l_{m}) = (1-\rho)\Big(1+\frac{\hat{\xi}(l_{m})}{\hat{\sigma}(l_{m})}\big(\phi_{l_{m}} -\hat{\mu}(l_{m})\big)\Big)^{-1/\hat{\xi}(l_{m})},
\end{equation}
where $\hat{\sigma}, \hat{\xi}$,  and $\hat{\mu}$ are estimates of  $\sigma, \xi$ and $\mu$, respectively, and   $\phi_{l_m} = f(\gamma_0)$. The expression in $\eqref{Outage_eq}$ accurately characterises the SINR distribution tail at the observed locations when $N$ is large. However, the predictive capabilities of $\eqref{Outage_eq}$ are limited to the locations with observations and will fail to model the outage probability at new locations. This is because features such as the number of scatters and line-of-sight (LOS) blockage, which strictly depend on the environment characteristics, may change even for movements in the order of half-wavelength, thus affecting the distribution of the SINR and its tail in particular. In this sense, predicting the parameters of the GPD at new locations based on their observations and exploring the spatial correlation of the environments seems a more effective strategy. 
\subsection{Radio-map-based GPD parameters prediction
}
For each BS $b$, the parameters of the GPD at unobserved locations $l_{m}'$ can be predicted using Gaussian processes \cite{perez2024evt}. Defining the sets of estimated parameters of the GPD at observed locations, we have $\hat{\mathbf{s}}_b(\mathcal{L}) = \big[\hat{\sigma}(l_{1}), \hat{\sigma}(l_{2}),...,\hat{\sigma}(l_{m})\big]^T$, $\hat{\mathbf{e}}_b(\mathcal{L}) = \big[\hat{\xi}(l_{1}), \hat{\xi}(l_{2}),...,\hat{\xi}(l_{m})\big]^T$, and $\hat{\mathbf{u}}_b(\mathcal{L}) = \big[\hat{\mu}(l_{1}), \hat{\mu}(l_{2}),...,\hat{\mu}(l_{m})\big]^T$. Then, the sets are normalised as
\begin{subequations}
\begin{align}  \Bar{\mathbf{s}}_b(\mathcal{L}) &= \frac{\hat{\mathbf{s}}_b(\mathcal{L})-\text{SM}\big(\hat{\mathbf{s}}_b(\mathcal{L})\big)}{\text{SSD}\big(\hat{\mathbf{s}}_b(\mathcal{L})\big)}\label{Normalization1},\\
    \Bar{\mathbf{e}}_b(\mathcal{L}) &= \frac{\hat{\mathbf{e}}_b(\mathcal{L})-\text{SM}\big(\hat{\mathbf{e}}_b(\mathcal{L})\big)}{\text{SSD}\big(\hat{\mathbf{e}}_b(\mathcal{L})\big)}\label{Normalization2},\\
    \Bar{\mathbf{u}}_b(\mathcal{L}) &= \frac{\hat{\mathbf{u}}_b(\mathcal{L})-\text{SM}\big(\hat{\mathbf{u}}_b(\mathcal{L})\big)}{\text{SSD}\big(\hat{\mathbf{u}}_b(\mathcal{L})\big)}\label{Normalization3},
\end{align}
\end{subequations}
where SM$(\cdot)$ and SSD$(\cdot)$ are functions computing the sample mean and sample standard deviation, respectively. The sets $ \Bar{\mathbf{s}}_b(\mathcal{L})$, $ \Bar{\mathbf{e}}_b(\mathcal{L})$ and  $\Bar{\mathbf{u}}_b(\mathcal{L})$ can be seen as noisy observations of multi-dimensional Gaussian processes \cite{williams2006gaussian}, \textit{i.e.,}
\begin{subequations}
\begin{align}\label{GP}
\Bar{\mathbf{s}}_b(\mathcal{L})&=\Bar{\Bar{\mathbf{s}}}_b(\mathcal{L})+\tilde{\mathbf{s}}_b,\\
\Bar{\mathbf{e}}_b(\mathcal{L})&=\Bar{\Bar{\mathbf{e}}}_b(\mathcal{L})+\tilde{\mathbf{e}}_b,\\
\Bar{\mathbf{u}}_b(\mathcal{L})&=\bar{\Bar{\mathbf{u}}}_b(\mathcal{L})+\tilde{\mathbf{u}}_b,
\end{align}
\end{subequations}
where $\bar{\bar{\mathbf{s}}}_b(\mathcal{L})\sim \mathcal{N}(\mathbf{0}, \mathbf{R}_{b,\scaleto{\mathcal{L}\mathcal{L}}{4pt}}^s)$, $\bar{\bar{\mathbf{e}}}_b(\mathcal{L})\sim \mathcal{N}(\mathbf{0}, \mathbf{R}_{b,\scaleto{\mathcal{L}\mathcal{L}}{4pt}}^e)$, and $\bar{\bar{\mathbf{u}}}_b(\mathcal{L})\sim \mathcal{N}(\mathbf{0}, \mathbf{R}_{b,\scaleto{\mathcal{L}\mathcal{L}}{4pt}}^u)$ are Gaussian processes. Moreover $\tilde{\mathbf{s}}_b\sim \mathcal{N}(\mathbf0, \lambda_s^2\mathbf{I})$, $\tilde{\mathbf{e}}_b\sim \mathcal{N}(\mathbf{0}, \lambda_e^2\mathbf{I})$, and $\tilde{\mathbf{u}}_b\sim \mathcal{N}(\mathbf{0}, \lambda_u^2\mathbf{I})$ represent the distribution of the observation noise. The entries of the correlation matrix $ \mathbf{R}_{b,\scaleto{\mathcal{L}\mathcal{L}}{4pt}}^u$ are well captured by the Gudmundson correlation model. In contrast, the Matérn model provides a more accurate representation for $\mathbf{R}_{b,\scaleto{\mathcal{L}\mathcal{L}}{4pt}}^s$ and  $\mathbf{R}_{b,\scaleto{\mathcal{L}\mathcal{L}}{4pt}}^e$ \cite{perez2024evt, kallehauge2022predictive}. The $(i,j)-$th entry of the correlation matrices following Gudmundson and Matérn models is respectively given by
\begin{align}
\mathbf{R}^{u}_{b,\scaleto{\mathcal{L}\mathcal{L}}{4pt}}(i,j) &= \omega_{u}^2\exp\Big(\frac{||l_{m,i}-l_{m,j}||}{r_{u}}\Big), \label{Covariance1}\\
\mathbf{R}^{s/e}_{b,\scaleto{\mathcal{L}\mathcal{L}}{4pt}}(i,j) \!& = \!\omega_{s/e}^2\frac{2^{1\!-\!\nu_{s/e}}}{\Gamma({\nu_{s/e}})}\!\Bigg(\!\!\frac{\sqrt{\nu_{s/e}}||l_{m,i}\!-\!l_{m,j}||}{r_{s/e}}\!\Bigg)^{\nu_{s/e}}\label{Covariance2}\\ \nonumber &\times B_{\nu_{s/e}}\!\Bigg(\!\!\frac{\sqrt{\nu_{s}}||l_{m,i}\!-\!l_{m,j}||}{r_{s/e}}\!\Bigg),\ \ \ i,j \in [1, N]
\end{align}
where $\omega_{u}^2, \omega_{s/e}^2, r_u,r_{s/e}$, and $\nu_{e/s}$ are hyperparameters that must be estimated from the data. 

To predict the parameters of the GPD at a set of \( M' \) unobserved locations \( \mathcal{L}'\), we assume that the joint distribution of the normalised observations, as defined in \eqref{Normalization1}, \eqref{Normalization2}, and \eqref{Normalization3}, is Gaussian across the coverage area of each BS \( b \), as these parameters are expected to vary smoothly over space.  This assumption, validated in~\cite{perez2024evt, kallehauge2022predictive}, allows us to compute the normalised predictive mean and covariance matrices of the parameters at the unobserved locations using standard Gaussian process regression techniques~\cite{williams2006gaussian}. The corresponding denormalised mean and covariance can then be obtained as
 
\begin{align}
\hat{\mathbf{d}}_b(\mathcal{L}') &= \text{SSD}\big(\mathbf{d}_b(\mathcal{L})\big)\times\nonumber
\\&\Big(\!\mathbf{R}^{d}_{b,\scaleto{\mathcal{L'}\mathcal{L}}{4pt}} (\mathbf{R}^{d}_{b,\scaleto{\mathcal{L}\mathcal{L}}{4pt}}\! + \!\omega^2_{d} \mathbf{I}_M)^{-1}\bar{\mathbf{d}}_b(\mathcal{L})\!\Big) \!+ \!\text{SM}\big(\hat{\mathbf{d}}_b(\mathcal{L})\big),\label{meanV}\\
\mathbf{R}^{d(\mathcal{L}')}_b& = \big [\text{SSD}\big(\hat{\mathbf{d}}_b(\mathcal{L})\big)\big]^2 \times \nonumber\\
&\Big(\!\mathbf{R}^{d}_{b,\scaleto{\mathcal{L'}\mathcal{L'}}{4pt}} - \mathbf{R}^{d}_{b,\scaleto{\mathcal{L'}\mathcal{L}}{4pt}} (\mathbf{R}^{d}_{b,\scaleto{\mathcal{L}\mathcal{L}}{4pt}} + \lambda^2_{d}\mathbf{I}_{M})^{-1} \mathbf{R}^{d}_{b,\scaleto{\mathcal{L}\mathcal{L}}{4pt}}\!\Big),\label{meanC}
\end{align}
where $d\leftarrow\{u,s,e\}$, $\hat{\mathbf{d}}_b~\leftarrow~\{\hat{\mathbf{u}}_b,\hat{\mathbf{s}}_b,\hat{\mathbf{e}}_b\}$, and $\bar{\mathbf{d}}_b~\leftarrow~\{\bar{\mathbf{u}}_b,\bar{\mathbf{s}}_b,\bar{\mathbf{e}}_b\}$. The Gaussian assumption, together with the hyperparameters obtained from \eqref{meanV} and \eqref{meanC}, enables a full statistical characterisation of the GPD parameter distribution at any spatial location. As in \cite{perez2024evt}, we assume that the predicted shape and scale parameters at new locations are given by the corresponding predictive means in \eqref{meanV}. This assumption is justified by the relatively small predictive variances associated with these parameters, which implies a negligible effect on the characterisation of the tail behaviour. In contrast, the predictive variances of the thresholds are substantially larger. To account for this uncertainty and to ensure robustness against prediction error, the thresholds are defined as empirical $\tau$-quantiles as

\begin{equation}\label{quantile} 
 \hat{\mu}^{\tau}(l_{m}') = Q^{-1}\Big(\tau, \hat{\mathbf{u}}_{m'}(\mathcal{L}'), \mathbf{R}_b^{\hat{u}(\mathcal{L}')}(m',m')\Big).   
\end{equation}
\subsection{Constraint reformulation and power scaling computation}
Given the results in \eqref{meanV}$-$\eqref{quantile}, the predictive outage probability at the unobserved location $l'_m$ is given by 
\begin{equation}\label{Outage_eq_3}
    \mathcal{O}(l_{m}') = (1-\rho)\Big(1+\frac{\hat{\xi}(l_{m}')}{\hat{\sigma}(l_{m}')}\big(\phi_{l_{m}'} -\hat{\mu}^{\tau}(l_{m}')\big)\Big)^{-1/\hat{\xi}(l_{m}')}.
\end{equation}
Notice that the estimated parameters of the GPD depend on the transmit power used when obtaining the SINR measurements. Let us assume that the SINR samples are normalised to a common power $p_0$; thus, finding the minimum power $p_{b, min}$, which ensures the outage demands at each spatial position, translates to finding an optimum power scaling factor $\kappa_b\in\mathbb{R}^+$ such that $p_{b, min} = \kappa_b\times p_0$. Notice that $\kappa_b$ introduces variations to the distribution of the SINR and, consequently, to the estimated scale, shape, and threshold parameters but the specific variations depend on the transformation function $f(\cdot)$. For instance, assuming $f = -\ln(\cdot)$ for the RV $Z$, letting $Z_t=-\ln(Z)$, and introducing a scaling $\kappa_b$, we have
\begin{equation}\label{transformation}
   f(\kappa_b Z) = -\ln(\kappa_b Z) = -\ln(\kappa_b) - \ln(Z) = -\ln(\kappa_b)+Z_t.
\end{equation}
Therefore, the scaling factor only introduces a shift of $-\ln(\kappa_b)$ to the RV $Z$, which updates the GPD parameters as follows
\begin{subequations}
\begin{align}
   \hat{\xi}(l_{m}') & \leftarrow \hat{\xi}(l_{m}'),\label{new_e}\\
    \hat{\sigma}(l_m') & \leftarrow \hat{\sigma}(l_m'),\label{new_s}\\
    \hat{\mu}^{\tau}(l_{m}') & \leftarrow \hat{\mu}^{\tau}(l_{m}') - \ln(\kappa_b)\label{new_u}. 
\end{align}
\end{subequations}
This is, shape and scale remain constant, while the threshold experiences a shift. Determining the minimum power that satisfies the outage constraint $\zeta$ reduces to solving for $\kappa_b$. This is achieved by substituting the updated GPD parameters from \eqref{new_e}$–$\eqref{new_u} into \eqref{Outage_eq_3} and subsequently isolating $\kappa_b$ which results in
\begin{equation}\label{kappa}
    \kappa_b = \gamma_0\exp\Bigg(\frac{\hat{\sigma}(l_{m}')}{\hat{\xi}(l_{m}')}\bigg[\Big(\frac{\zeta}{1-\rho}\Big)^{-\hat{\xi}(l_{m}')}-1\bigg]+\hat{\mu}^{\tau}(l_{m}')\Bigg).
\end{equation}
\section{Radio map fine-tuning and BS selection policy}\label{Section4}
Following the results in \eqref{kappa}, maps of minimum transmit power can be constructed for each BS. However, such maps require further processing and a policy defined for the BS selection in the HO processes as discussed next.
\subsection{Power map creation and filtering}
For each BS $b$ and position $l_{m}'$ in the map, we can determine the minimum required power for a given target $\zeta$ as $p_{b,min} = \kappa_b \times p_0$ as previously discussed. Once this step is done, we obtain $B$ power maps $\mathcal{P}_1, \mathcal{P}_2,..., \mathcal{P}_B$. The maps $\mathcal{P}_b$ are optimal for static UEs, which require a constant transmit power for a given outage probability target at a particular location. However, for moving UEs, the transitions in the power transmission from one position in the map to another nearby position might be abrupt. To ensure smoother transitions between adjacent positions addressing spatial variations in power levels, and guarantee more efficient BS selection, it is recommended to somewhat filter the high-frequency components in the obtained maps $\{\mathcal{P}_b\}$. In this sense, techniques used for digital image processing come in handy. 

Applying a two-dimensional filter $F(\mathcal{P}_b,\theta)$ to a power map $\mathcal{P}_b$ results in new maps as
\begin{equation}\label{Filter}
    \mathcal{P}_{b,filt} = F(\mathcal{P}_b, \theta) = G(\theta)*\mathcal{P}_b\ \forall b,
\end{equation}
where $\theta$ is a hyperparameter of the chosen kernel $G(\cdot)$. For instance, when applying a Gaussian kernel $G(\theta)$ to a power map $\mathcal{P}_b$, the convolution averages the pixels (herein power levels) based on their spatial closeness and their value similarity. The Gaussian kernel is chosen due to its smooth weighting of neighbouring values, which preserves spatial continuity and reduces abrupt transitions. Its isotropic nature ensures that nearby power levels contribute more significantly, making it especially suitable for smoothing spatially varying maps. In that case, the new power value at coordinates (x,y) is given by the following operation
\begin{equation}\label{conv}
    \mathcal{P}_{b,filt}(\text{x,y})= \sum_{i=-C_{\text{x}}}^{C_{\text{x}}}\sum_{j=-C_{\text{y}}}^{C_{\text{y}}}G(i,j;\theta)\times \mathcal{P}_b(\text{x}-i,\text{y}-j),
\end{equation}
where $C_{\text{x}}$ and $C_{\text{y}}$ represent the Kernel radius in x and y axes, respectively. A suitable design criterion is to set $C_{\text{x}} = C_{\text{y}} = \lceil 2\theta \rceil$ as this ensures that the Gaussian kernel includes more than $95\%$ of its total volume within two standard deviations $\theta$. The Kernel is given by
\begin{equation}\label{kernel}
    G(m,n; \theta) = \frac{1}{2\pi\theta^2}\exp\bigg(-\frac{m^2+n^2}{2\theta^2}\bigg).
\end{equation}

A higher $\theta$ results in more blurring, as the kernel has a broader range and influences more neighboring power values \cite{gonzalez2009digital}. However, because of the way the filter works, certain power levels in the new maps $\mathcal{P}_{b,filt}$ may be below those in $\mathcal{P}_b$ at the same locations. This issue potentially affects the system's performance at those locations, thus causing violations of the outage probability target $\zeta$. To avoid this problem, we perform a peak detection such that only the maximum power between the predicted and filtered maps is chosen at each spatial position, \textit{i.e.,}
\begin{equation}\label{finalpower}
    \mathcal{P}_{b}^\circ = \max(\mathcal{P}_{b,filt},\mathcal{P}_b).
\end{equation}

Each map $\mathcal{P}_b$ represents the minimum transmit power required by BS $b$ to satisfy the outage constraint at every location. Given these maps, the next step is to determine which BS should be active in serving a UE located at position $l_m'$ at time instant $T_i$. This decision is governed by the activation function $\mathbf{a}(\cdot)$, whose design is described next.

\begin{algorithm}[t!]
\caption{Filtered-minimum-power-map generation for URC}
\hspace*{\algorithmicindent} \textbf{Inputs:} $\{\Gamma_b\}$, $\rho$,   $\mathcal{L}$, $
\mathcal{L{'}}$, $\zeta$, $\tau$, $p_0$, $\gamma_0$, $\theta$  \\
\hspace*{\algorithmicindent}
\textbf{Outputs:} $\{\mathcal{P}_{b}^\circ\}$
\begin{algorithmic}[1]
\State Compute $\hat{\mu}$ with \eqref{threshold_eq}
\State Compute $\psi(l_{m})$ with \eqref{excess}
\State Obtain log-likelihood estimates $\hat{\sigma}$ and $\hat{\xi}$
\State Obtain normalised sets $\bar{\mathbf{s}}(\mathcal{L}), \bar{\mathbf{e}}(\mathcal{L}) $ and $ \bar{\mathbf{u}}(\mathcal{L})$ with \eqref{Normalization1}, \eqref{Normalization2} and \eqref{Normalization3}
\State Obtain log-likelihood estimates of  $\omega_u^2$,  $\omega_{s}^2$, $\omega_{e}^2$, $r_u$, $r_s$, $r_e$ $\nu_s$, $\nu_e$, $\lambda_{u}^2$, $\lambda_{s}^2$ and $\lambda_{e}^2$ 
\State Compute the correlation matrices with \eqref{Covariance1} and \eqref{Covariance2}
\State Compute the denormalised predictive mean and covariance with \eqref{meanV} and \eqref{meanC} for $\mathcal{L}'$
\State For each entry of $\hat{\mathbf{u}}(\mathcal{L'})$ in \eqref{meanV} compute the $\tau-$quantile $\hat{\mu}^{\tau}(l_{m}')$  with \eqref{quantile}
\State Compute $\kappa_b$ $\forall l_{m}'\in\mathcal{L}'$, $\forall b$ with \eqref{kappa}
\State Compute $p_{b,min}= \kappa_b\times p_0$ $\forall l_{m}'\in\mathcal{L}'$  $\forall b$ to obtain $\mathcal{P}_b$
\State Compute the filtered maps $\mathcal{P}_{b,filt}$ $\forall b$ with \eqref{Filter}
\State Compute the minimum power maps $\mathcal{P}_{b}^\circ \ \forall b$ with \eqref{finalpower}
\end{algorithmic}
\end{algorithm}

\begin{algorithm}[t!]
\caption{Filtered-minimum-power-map-based HO decision for URC}
\hspace*{\algorithmicindent} \textbf{Inputs:} $\{\mathcal{P}_{b}^\circ\}$, $\Delta p$, $\Delta T$, $p_{max}$ \\
\hspace*{\algorithmicindent}
\textbf{Outputs:} $b^*, p_{b^*,min}$
\begin{algorithmic}[1]
\If{there is DL data for a UE in location $l_m'$}
\State Initialise $p_{b^*,min}$ and $b^*$ with \eqref{BSselection}, $T_0=T_i$, $\mathbf{a}(b^*) = 1$
\While{DL channel to UE is active for transmission}
\State Update $T_i$ and $l_{m}'$
\If{$\mathcal{P}_{b*}^\circ(l_{m}')>p_{max}$  \textbf{or} \Big($T_i-T_0\ge\Delta T$ \textbf{and} $\min\big(\mathcal{P}_{1}^\circ(l_{m}'),...,\mathcal{P}_{B}^\circ(l_{m}')\big)-\mathcal{P}_{b*}^\circ(l_{m}')\ge \Delta p \Big )$}
\State Update $p_{b^*,min}$ and $b^*$ with 
\eqref{BSselection}, $T_0=T_i$,
\hspace*{\algorithmicindent}\hspace*{\algorithmicindent} \hspace*{\algorithmicindent}$\mathbf{a}(b^*) = 1$
\EndIf
\EndWhile
\EndIf
\end{algorithmic}
\end{algorithm}
\subsection{BS selection}
Clearly, the minimum values of the power vector $\mathbf{p}(l_{m}')$ in \textbf{P1} that guarantee the outage requirements for each BS $b$ at location $l_{m}'$ can be obtained from each map as $\mathcal{P}_{b}^\circ(l_{m}'$). If the activation function $\mathbf{a}$ in 
\textbf{P1} only depended on the position $l_{m}'$, the active BS  $b^*$ would be selected as the one with minimum power at that location. 
However, for practical application with moving UEs, $\mathbf{a}$ must also include a time variable $\Delta T$ (hold timer) that controls the frequency of the HOs, potentially mitigating the ping-pong effect between BSs. Contrary to the typical values of the TTT, $\Delta T$ may even be in the order of units of seconds since the HO decisions are based on the UE localisation instead of the SINR or RSS monitoring. Inserting a hysteresis margin $\Delta p$  also seems reasonable since an HO to a new BS does not represent significant energy saving if the power difference is nearly zero. 

Summarising, an HO to a BS $b^*$ is triggered if the time since the last HO is equal or larger than the hold timer margin, \textit{i.e.,} $\Delta T \le T_i - T_0$, where $T_i$ and $T_0$ depict the current time instant and the time instant of the last HO, respectively. The HO will be triggered if the chosen BS $b^*$ offers a power saving equal to or larger than $\Delta p$. Exceptionally, a new HO decision can be issued if the active BS  $b^*$ reaches its peak power value (potential maximum power violation and outage probability degradation), independently of the controls offered by $\Delta T$ and $\Delta p$.  When an HO is triggered at time $T_i$ and location $l_m'$, the transmit power $p_{b^*\!,min}$ and the number of the serving BS $b^*$ are determined as
\begin{equation}\label{BSselection}
[p_{b^*\!,min}, b^*]=\min\big(\mathcal{P}_{1}^\circ(l_{m}'),...,\mathcal{P}_{B}^\circ(l_{m}')\big),
\end{equation}
thus, the activation function $\mathbf{a}(\cdot)$ gets `1' in position $b^*$ at time $T_i$ and location $l_m'$. 

\textbf{Algorithm~1} and \textbf{Algorithm~2} capture the previously described procedures for power map construction and HO decision-making, respectively, which together constitute the solution to \textbf{P1}. Note that \textbf{Algorithm~1} is executed once to generate the maps, whereas \textbf{Algorithm~2} operates continuously while the BSs are actively serving UEs.

\begin{table}[t!]
    \centering
    \caption{Simulation parameters}
    \label{table_1}
    \begin{tabular}{l l}
        \hline
        \textbf{Parameter} & \textbf{Value} \\
            \hline             
            $p_0/p_{b'}$ & 0/0 dBm \cite{kallehauge2022predictive}\\
            $p_{max}$ & 50 dBm\\ 
            $\gamma_0$ & 10 dB \\
            $B/B'$ & 4/2\\
            $N$  & $10^5$ \\
            $NF$ &  7 dB\\
            $BW$ & $1$ MHz \\
            Frequency & 2.5 GHz \\
            $\tau$ &  $10^{-2}$\\
            $\rho$ & $0.99$ \cite{dumouchel1983estimating}\\ 
             $\Delta p$ & \{0, 2, 3, 6\} dB\\
            $\Delta T$ & $0-10$ s\\ 
         $\zeta$ &  $10^{-5}, 10^{-3}$ \\  
         $\theta$ & 1\\
          $M$ &  $10^3$ \\ 
           BS $b/b'$ height &  6/10 m \cite{quadriga}\\  
            UE height &  1.5 m \cite{quadriga}\\ 
            UE speed & $\{10, 20\}$ km/h\\
            TTT & 36 ms\\    
            $G(\cdot)$ & Gaussian Kernel\\
           $f(\cdot)$ & $-\ln(\cdot)$\\
            number of multipath  clusters & 10\\
            \hline
    \end{tabular}
\end{table}
\section{Numerical Results}\label{section5}

To evaluate the performance of the proposed algorithms for solving \textbf{P1}, we consider an indoor factory with dimensions of 200 m $\times$ 130 m (26000 \text{m}$^2$) with 4 BSs deployed at coordinates (-95,60) m (BS 1), (-95,-60) m (BS 2), (95,60) m (BS 3), and (95,-60) m (BS 4), being (0,0) m the centre of the coordinate system\footnote{It should be noted that the approach presented in this work can also be applied to other scenarios such as urban or suburban environments, provided that the necessary data are available, for example multiple SINR samples across different locations and UE localization information.}. The region is divided into a $150\times 150$ grid (22500 locations), therefore with a vertical/horizontal spacing of 0.87/1.33 m. We assume 2 interfering BSs ($B' = 2$) outside the factory with coordinates $(0,400)$ m and $(0,-400)$ m. Each BS $b$ knows $N = 10^5$ SINR measurements at $M = 10^3$ different locations. The carrier frequency is set to 2.5 GHz, $p_0 = 0$ $\mathrm{dBm}$, and $p_{max} = 50$ $\mathrm{dBm}$. The noise power is given by $\upsilon^2 = -173.8+10\log_{10}BW+ NF$, with $NF$ (7~$\mathrm{dB}$) as the noise figure and  $BW$ (1 MHz) as the bandwidth. The environment and channel coefficients are simulated using QuaDRiGa v2.6.1 with the scenario  \text{``3GPP\_38.901\_InF\_LOS} for the indoor factory and  \text{``3GPP\_3D\_UMi\_NLOS"} for the channels from the interfering BSs\cite{quadriga}. We adopt the transformation function $f = -\ln(\cdot)$. All the simulation parameters are displayed in Table \ref{table_1}. 

We consider a UE that follows two different trajectories (See Fig. \ref{Combined_maps_power}). These are: 
\begin{itemize}
    \item The UE departs from coordinates (-85,70) m and moves at a constant speed of 20 km/h with a trajectory of 170 m (East) $\rightarrow$ 40 m (South)  $\rightarrow$ 170 m (West) $\rightarrow$ 40 m (South) $\rightarrow$ 170 m (East) $\rightarrow$ 40 m (South) $\rightarrow$ 170 m (West),
    \item The UE departs from coordinates (90,50) m and moves at a constant speed of 10 km/h with a trajectory of 85 m (West) $\rightarrow$ 80 m (South)  $\rightarrow$ 85 m (West)
\end{itemize}

Notice that in indoor factory scenarios, autonomous guided vehicles or robots typically move at low speeds. Under these conditions, the Doppler shift and spread are negligible due to the moderate speeds at the chosen carrier frequency.

\begin{figure}[t!]
    \centering
    \includegraphics[width = \columnwidth]{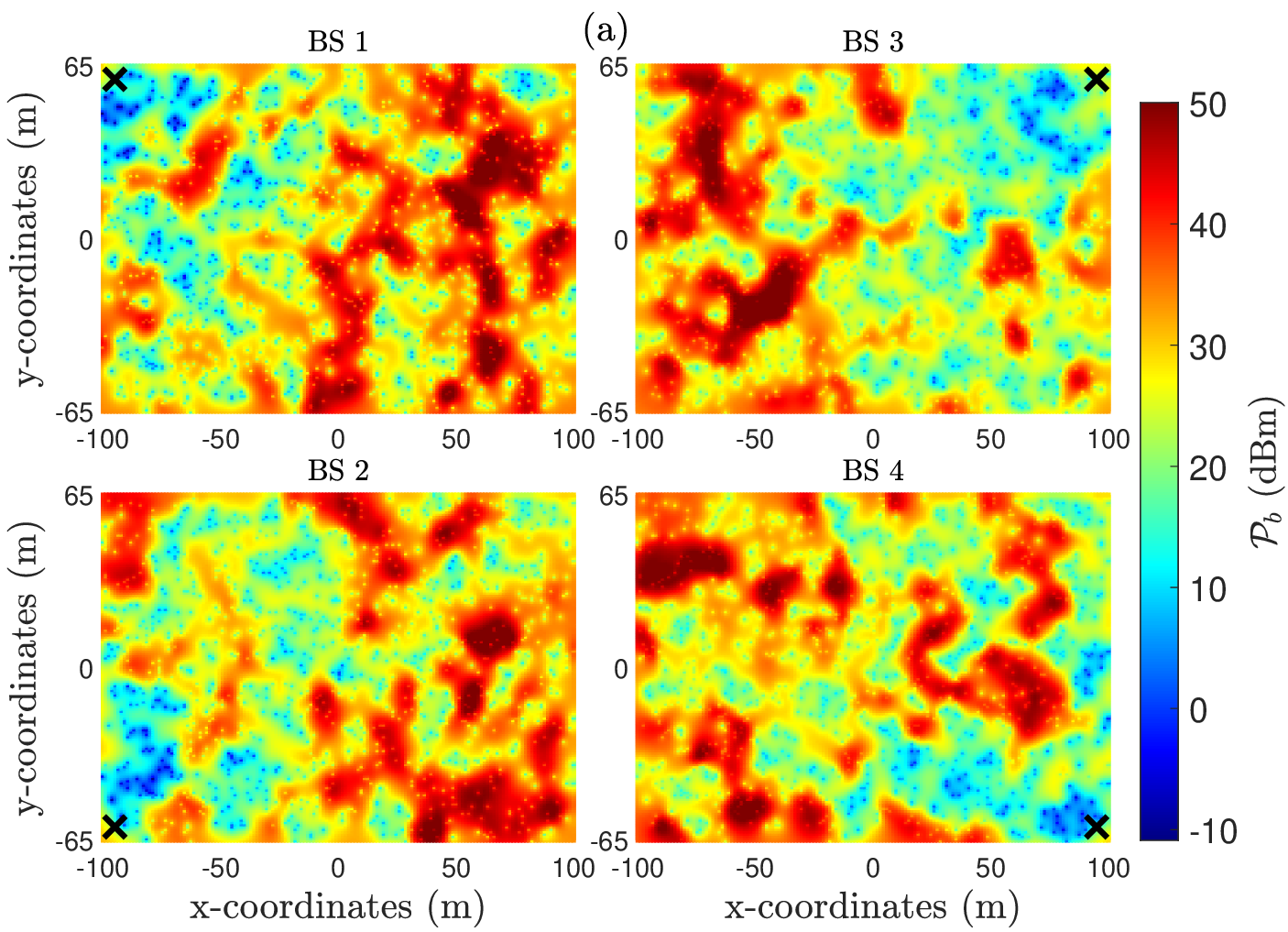}
    \includegraphics[width = \columnwidth]{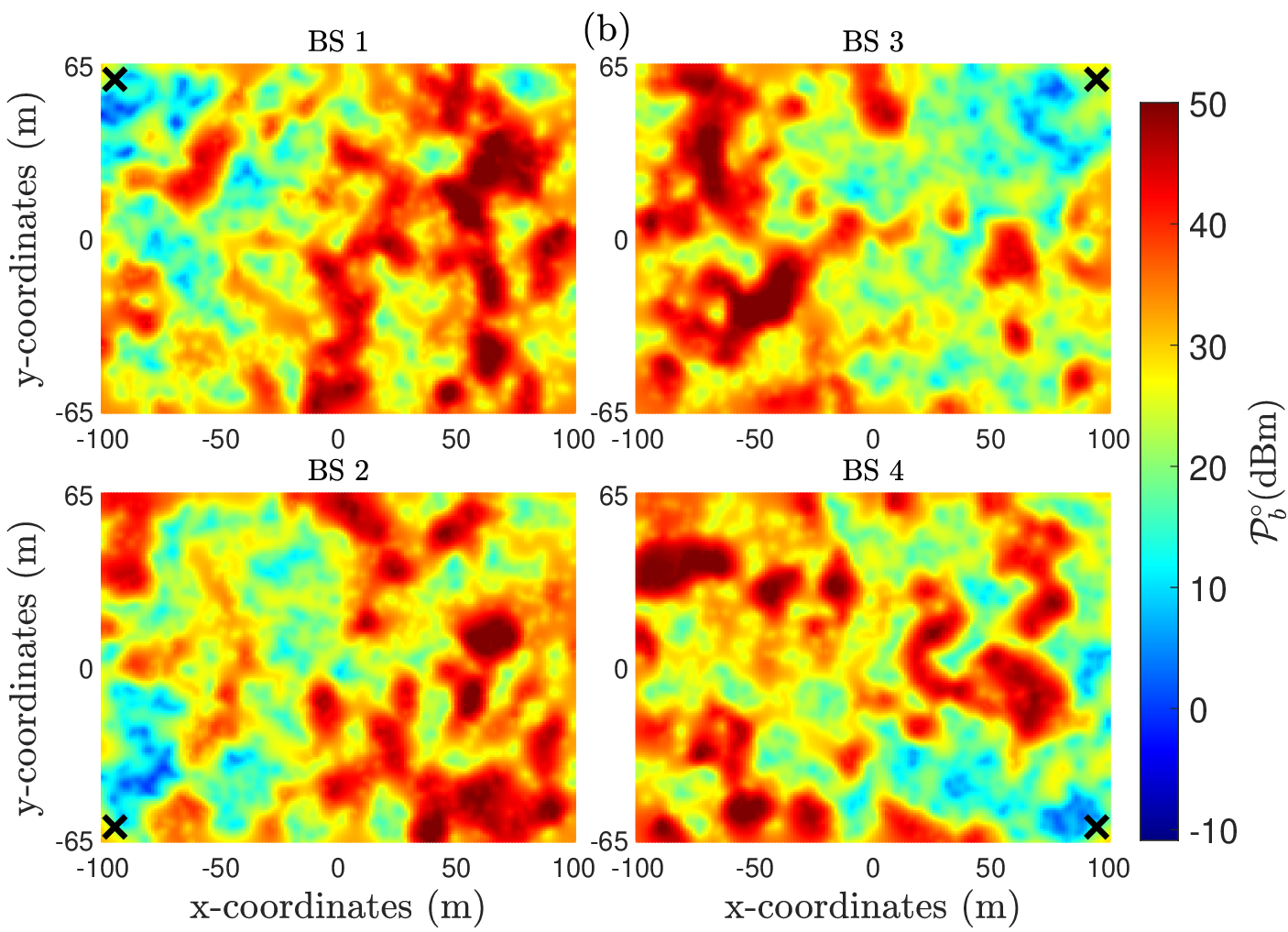}
    \caption{Predictive transmit power without filtering (a) and with filtering (b) $(\theta = 1)$, and $\zeta = 10^{-3}$ with \textbf{Algorithm 1}.}
    \label{predPower}
\end{figure}

Fig. \ref{predPower} shows the predictive transmit power for each BS at each location. Specifically, Fig. \ref{predPower} (a) shows the maps of transmit power without map filtering. Notice that the larger power values are allocated to the regions with poorer signal quality and do not have a uniform dependence on the distance from the BSs. Also, note that the radio map images are noisy, which implies significant changes in the allocated power with small movements of the UEs. Nevertheless, this is an expected feature of the system since variations in the order of half-wavelengths in the UE's position can cause considerable SINR degradation. These maps contain the optimal power allocations from each BS for static UEs. Fig. \ref{predPower} (b) displays the filtered power maps after removing the power map noise. Notice that the power transitions between adjacent points are smoother than the maps in Fig. \ref{predPower} (a).  Moreover, this significantly impacts the BS selections as we further discuss.

\begin{figure}[t!]
    \centering
    \includegraphics[width = \columnwidth]{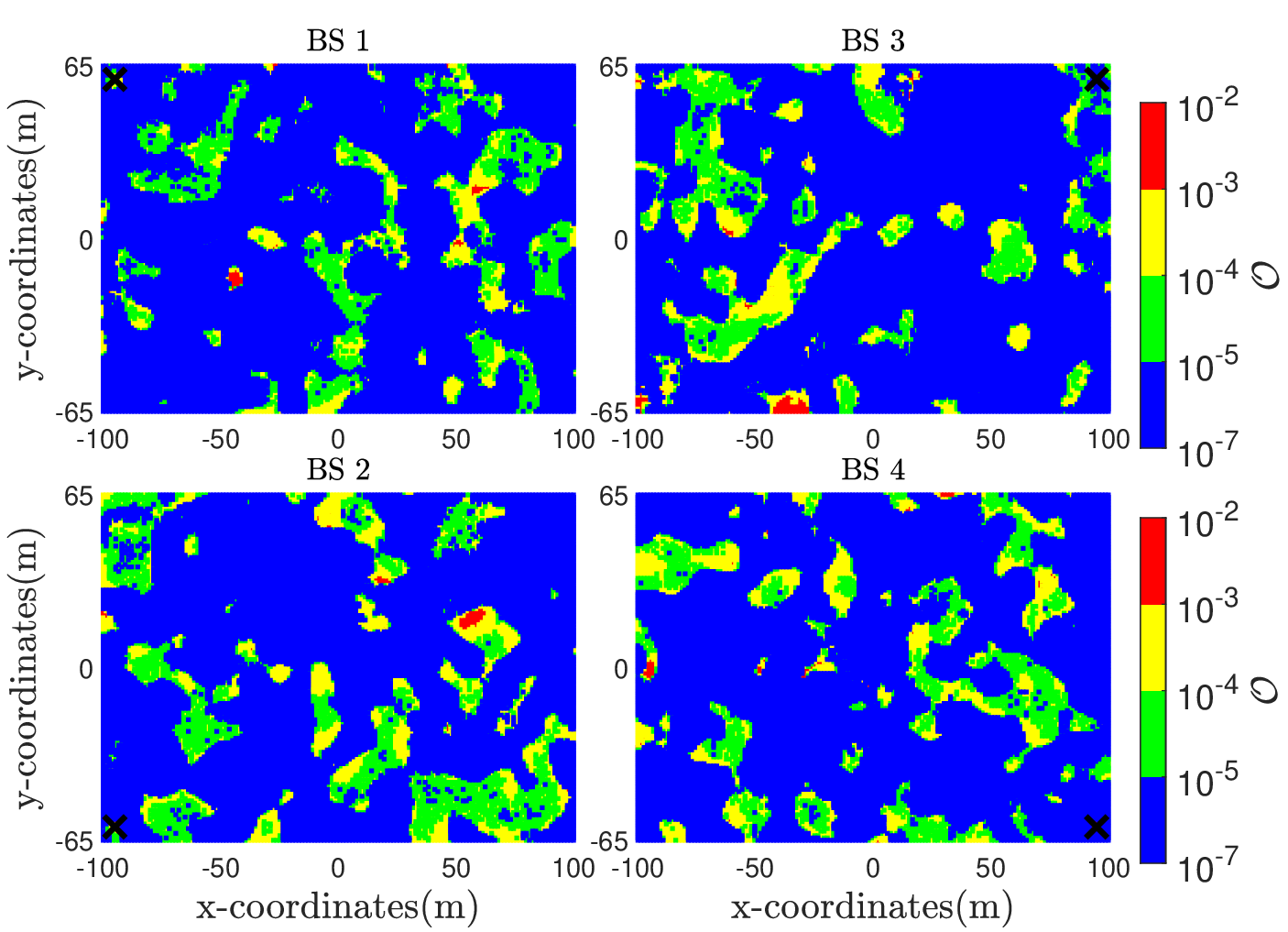}
    \caption{Actual outage probabilities for $\zeta = 10^{-3}$ at each BS $b$, obtained using the maps generated by \textbf{Algorithm~1}. The crosses represent the positions of the BSs}
    \label{RealOutage}
\end{figure}

Fig. \ref{RealOutage} displays the achieved outage probabilities with the proposed power allocation for each BS $b$. The target $\zeta = 10^{-3}$ is met in 99.59$\%$, 99.46$\%$, 99.23$\%$, and 99.45$\%$  of the coverage area (availability) for BSs $1-4$, respectively. Notice that the achieved outage probabilities are closer to the target $\zeta$ in the areas of poor signal quality, which means that the predictions are more accurate in these zones. This is a desirable behaviour since the areas of poor SINR conditions are more prone to outages, and therefore, a more accurate resource allocation is required.

\begin{figure}[t!]
    \centering
    \includegraphics[width = \columnwidth]{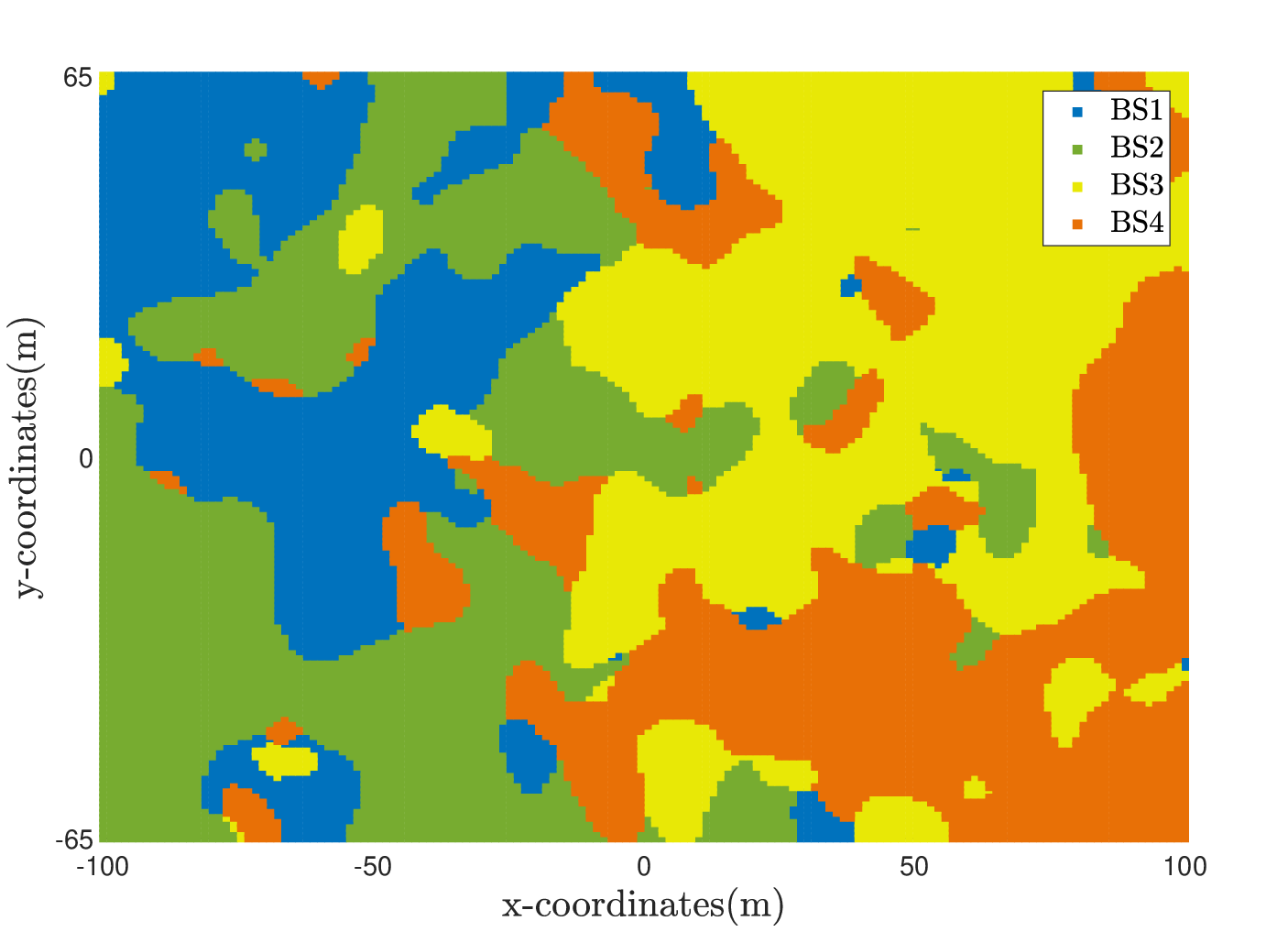}
    \caption{BS selection for minimum transmit power with filtering ($\theta = 1$). The selection is based only on the minimum power of the filtered maps at each spatial position generated by \textbf{Algorithm~1}.}
    \label{BS_selection}
\end{figure}
Fig. \ref{BS_selection} shows the optimal BS selection for each coordinate in the coverage area utilising the filtered power maps shown in Fig. \ref{predPower} (b). The application of filtering significantly alters the BS selection pattern, notably reducing the number of HOs as the UEs traverse specific paths. This reduction is key in reducing the ping-pong effect and enhancing the overall efficiency of the system by leveraging the control provided by $\Delta T$. This modification highlights the importance of integrating filtering strategies in dynamic BS selection processes to improve network stability and UE connectivity. Notice that the BS selection in the figure is scenario-specific, meaning that it depends on the particular environment and the locations of the BSs. As a result, the BS selection may change if the modelled coverage area is modified or if the macro-scale characteristics of the environment are altered, \textit{e.g.}, changes in scatterer positions or the presence of LOS blockages.

\begin{figure}[t!]
    \centering
    \includegraphics[width = \columnwidth]{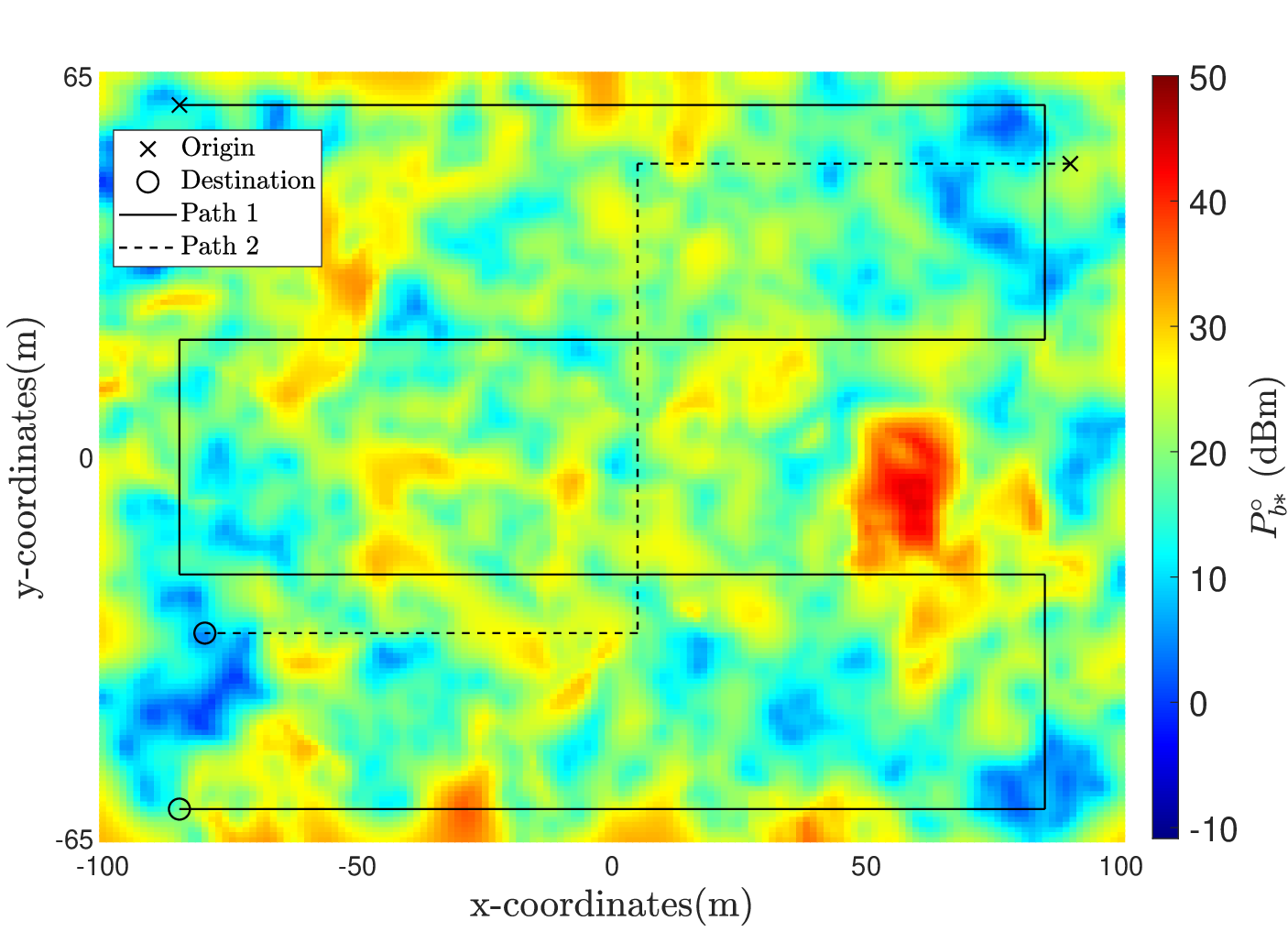}
    \caption{Minimum transmit power map with the BS selection in Fig. \ref{BS_selection} (b) at each spatial position, $\theta = 1$ and $\zeta = 10^{-3}$. The figure also shows the UE's trajectories moving at 20  km/h (Path 1) and 10  km/h (Path 2).}
    \label{Combined_maps_power}
\end{figure}

Fig. \ref{Combined_maps_power} shows the minimum transmit power in the shared coverage area using the BS selection policy with the map filtering from Fig. \ref{BS_selection} (b). Notice that the power levels are significantly reduced concerning the transmission from a single BS in Fig. \ref{predPower} (a) and Fig. \ref{predPower} (b). Moreover, the figure illustrates the two trajectories within the coverage area: Path 1 with a constant speed of 20 km/h and Path 2 with a constant speed of 10 km/h. Note that the number of possible trajectories is infinite in practice, and these two are only used as study cases to generate the results we discuss next. 

\begin{figure}[t!]
    \centering
    \includegraphics[width = \columnwidth]{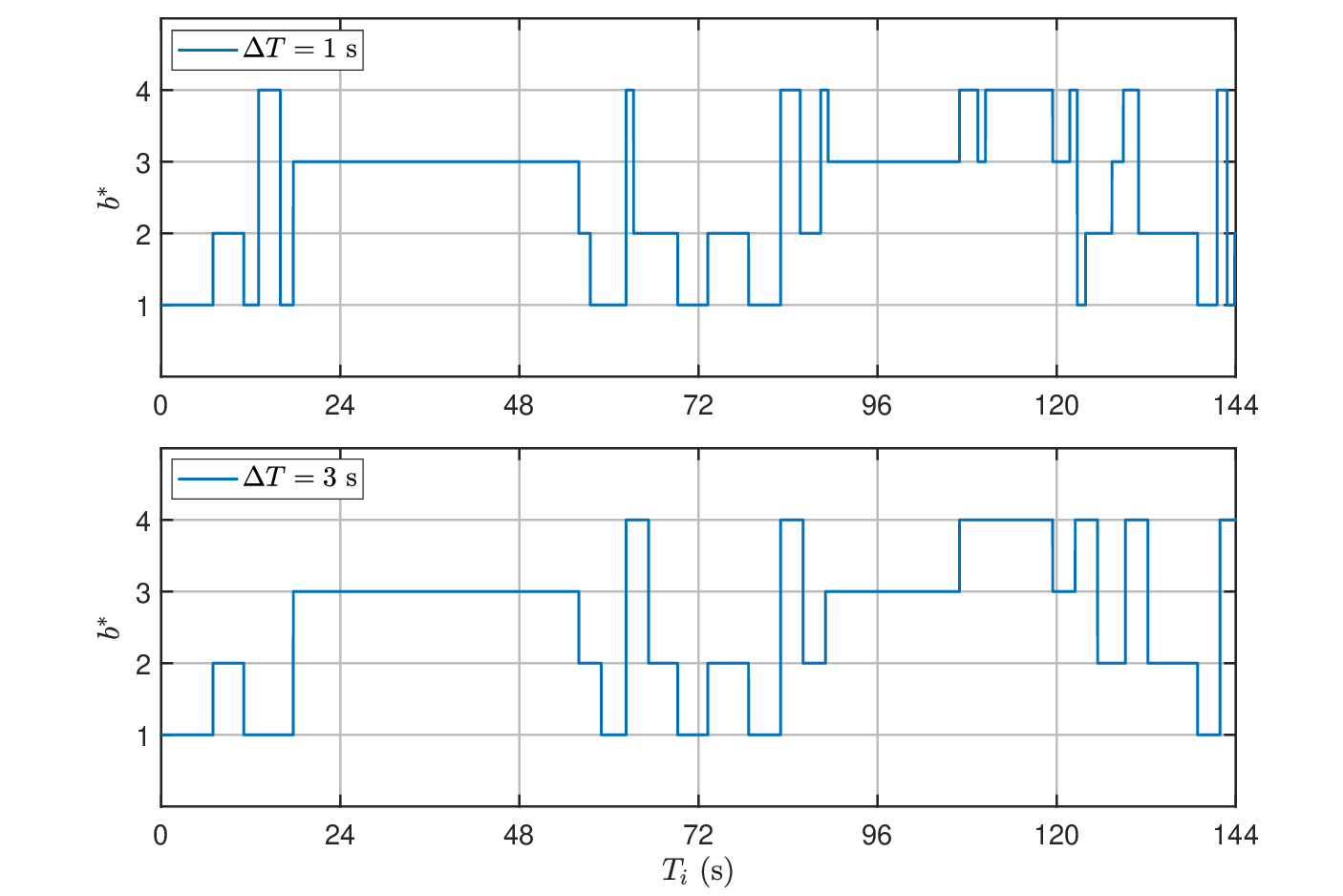}
    \caption{HO decisions along Path~1 using the proposed method with $\Delta p = 3$~dB, $\theta = 1$, and $\zeta = 10^{-3}$. \textbf{Algorithm~1} is executed once to construct the maps, while \textbf{Algorithm~2} runs continuously while the UE is being served.}
    \label{switchinghys_1}
\end{figure}

Fig. \ref{switchinghys_1} displays the BS selection for a UE travelling along the trajectory 1 displayed in Fig. \ref{Combined_maps_power}, utilising \textbf{Algorithm 2} with time steps of $\Delta T = 1$ s (top)  and $\Delta T = 3$ s (bottom). The average time between HOs is 4.8 s ($\Delta T = 1$ s) and 6.9 s ($\Delta T = 3$ s), translating to an HO every 26.69 m and 38.36 m, respectively. However, HOs may occur at shorter intervals near the limits set by $\Delta T$. This increased frequency is due to the UE encountering critical channel conditions that compromise the performance of the current serving BS or require a much higher transmit power, necessitating a switch to another BS with better channel conditions and lower power consumption. Note that figures similar to Fig.~\ref{switchinghys_1} can be constructed for any trajectory within the considered coverage area by following the BS selection policy illustrated in Fig.~\ref{BS_selection}, while adhering to the constraints imposed by $\Delta T$, $\Delta p$, and the maximum transmit power violation threshold. 
\begin{figure}[t!]
    \centering
    \includegraphics[width = \columnwidth]{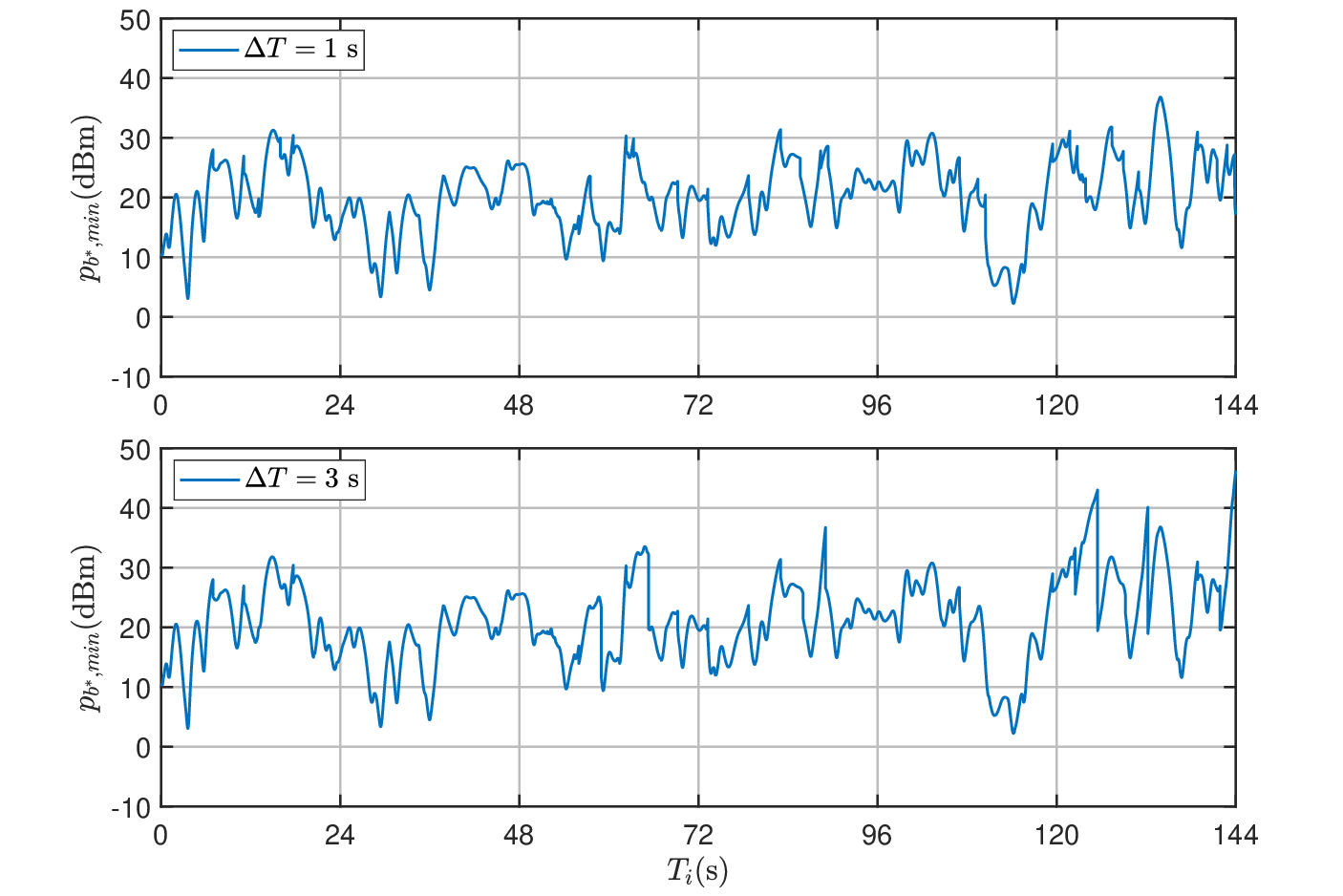}
    \caption{Transmit power along Path~1 using the proposed method with $\Delta p = 3$~dB, $\theta = 1$, and $\zeta = 10^{-3}$. \textbf{Algorithm~1} is executed once to construct the maps, while \textbf{Algorithm~2} runs continuously while the UE is being served.}
    \label{optimalpow1}
\end{figure}

\begin{figure}[t!]
    \centering
    \includegraphics[width = \columnwidth]{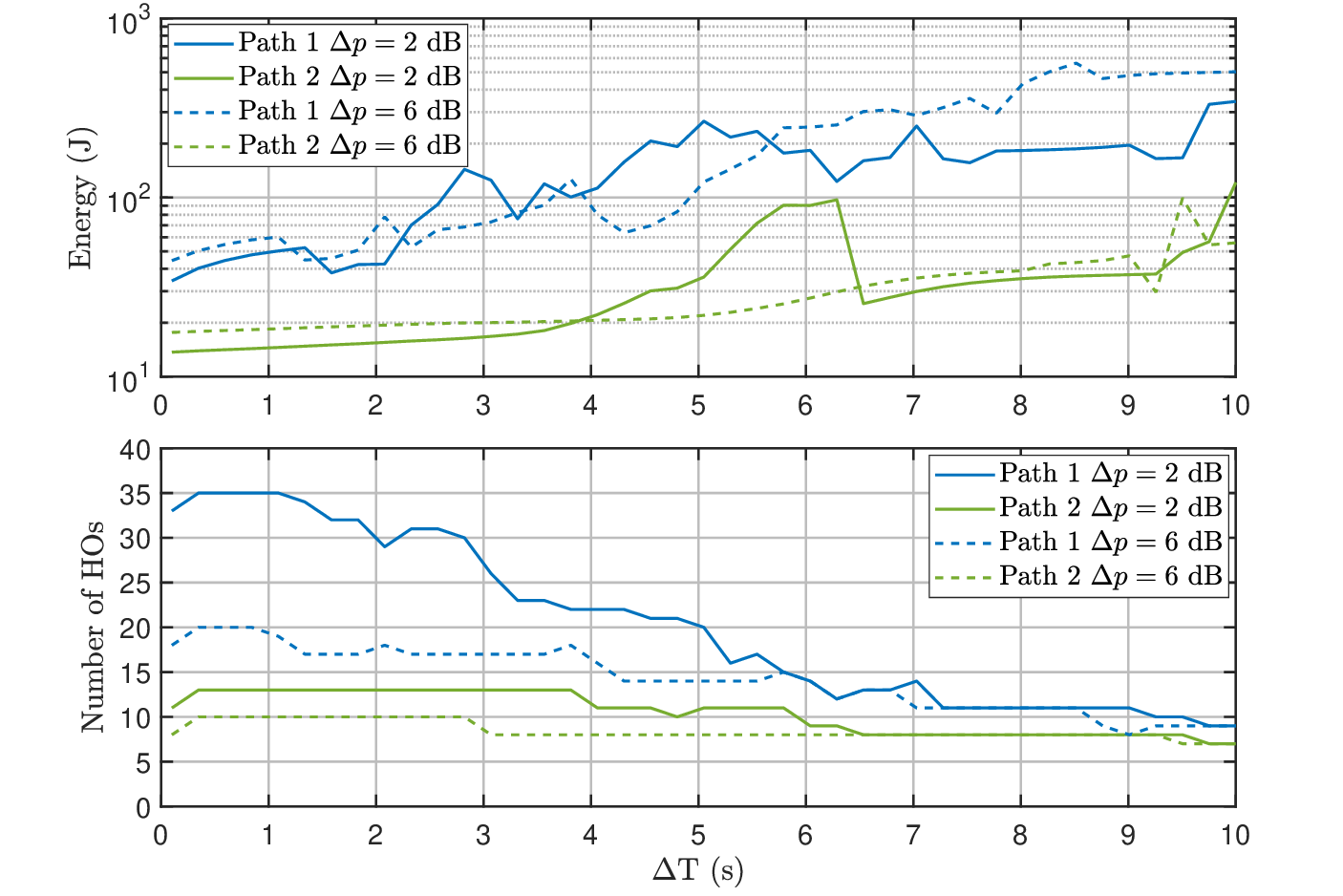}
    \caption{Energy consumption (top) and number of HOs (bottom) for different values of $\Delta T$ and $\Delta p$. We adopt $\theta =1$ and $\zeta= 10^{-3}$.}
    \label{energy_vs_deltat}
\end{figure}

\begin{figure}[t!]
    \centering
    \includegraphics[width = \columnwidth]{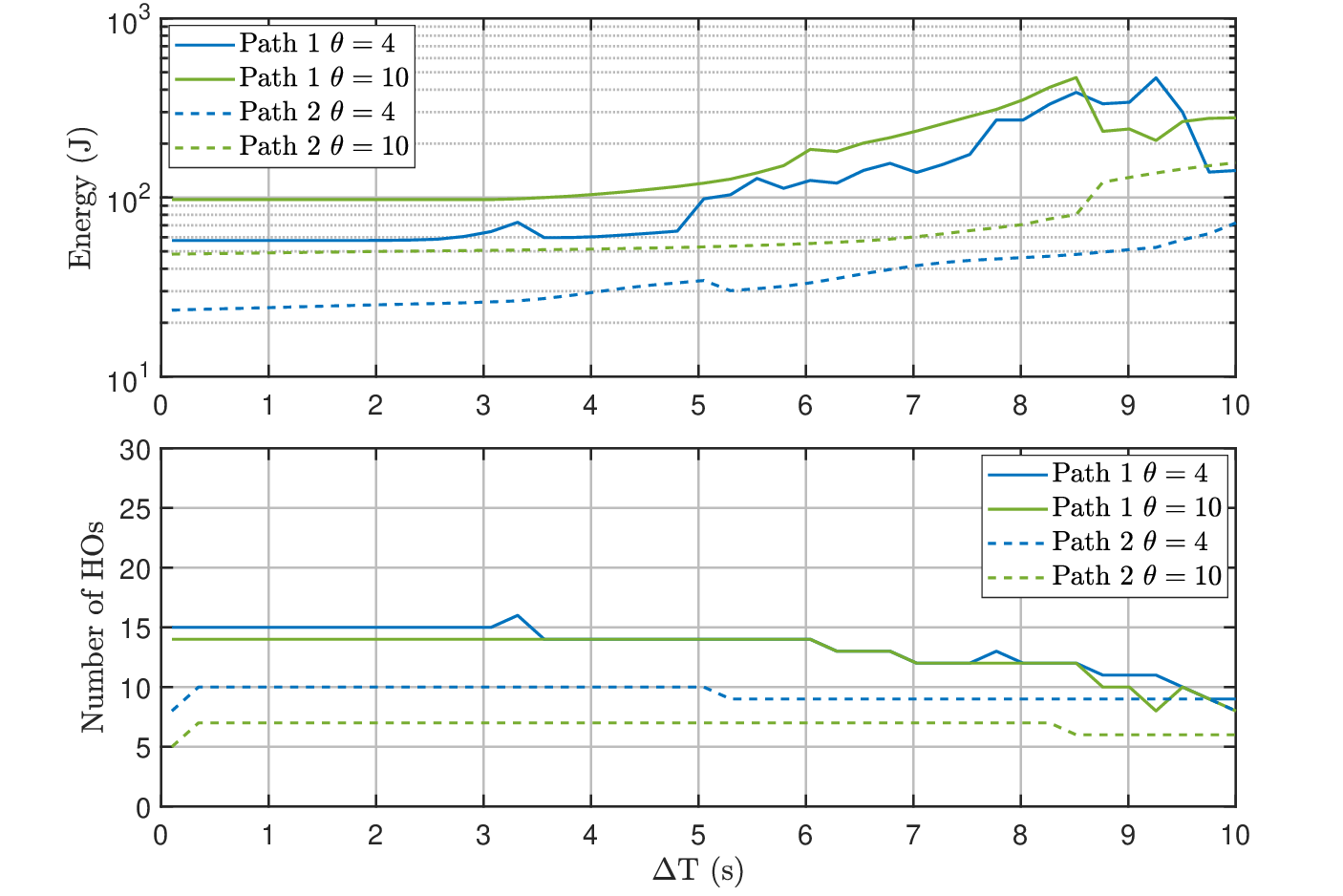}
    \caption{Energy consumption (top) and number of HOs (bottom) for different values of $\Delta T$ and $\theta$. We adopt $\Delta p = 3$ dB and $\zeta= 10^{-3}$.}
    \label{energy_vs_deltat2}
\end{figure}

Fig. \ref{optimalpow1} illustrates the transmission power along the path 1 for $\Delta T = 1$ s and $\Delta T = 3$ s, which is adjusted to meet the reliability requirements from the selected BS $b^*$ at the specific time $T_i$. It can be observed that the transmission power is dynamically adjusted, alternating between low and high levels. These fluctuations are correlated with the changing channel conditions. Specifically, low power is utilised in areas with a high likelihood of good signal reception. In contrast, high power is necessary in zones where the channel conditions are expected to be poor. Notice that the mean power values for $\Delta T = 1 \, \mathrm{s}$ and $\Delta T = 3 \, \mathrm{s}$ are $23.93 \, \mathrm{dBm}$ and $27.29 \, \mathrm{dBm}$, respectively. This indicates that the reduction in the number of HOs observed in Fig.~\ref{switchinghys_1} comes at the cost of an approximately $3.4$ $\mathrm{dB}$ increase in power.

Fig.~\ref{energy_vs_deltat} shows the energy consumption (top) and the number of HOs (bottom) for different values of $\Delta T$ and $\Delta p$. The energy is calculated as the product of transmit power and the time interval between trajectory snapshots (9 ms). The energy consumption increases with $\Delta T$ because switching to a new BS is delayed, even when the current BS is no longer the one with the minimum transmit power. Similarly, increasing $\Delta p$ makes HO decisions more restrictive from a power-saving perspective. Both parameters significantly reduce the number of HOs, meaning their selection should balance the trade-off between energy savings and minimising the number of HOs, depending on the primary design criteria.

Figure \ref{energy_vs_deltat2} illustrates the relationship between energy consumption (top) and the number of HOs (bottom) with varying values of $\theta$ and $\Delta T$. Higher values of $\theta$ not only increase transmission power but also influence how the transmit power transitions between adjacent positions. Specifically, larger $\theta$ values result in smoother transitions due to the broader filtering effect, which expands the area of influence for each power level at a given location. Consequently, the number of HOs tends to decrease as well as the ping-pong effect, as abrupt shifts to extremely low power levels in adjacent positions are less likely when a location already has high power level requirements.

\begin{figure}[t!]
    \centering
    \includegraphics[width = \columnwidth]{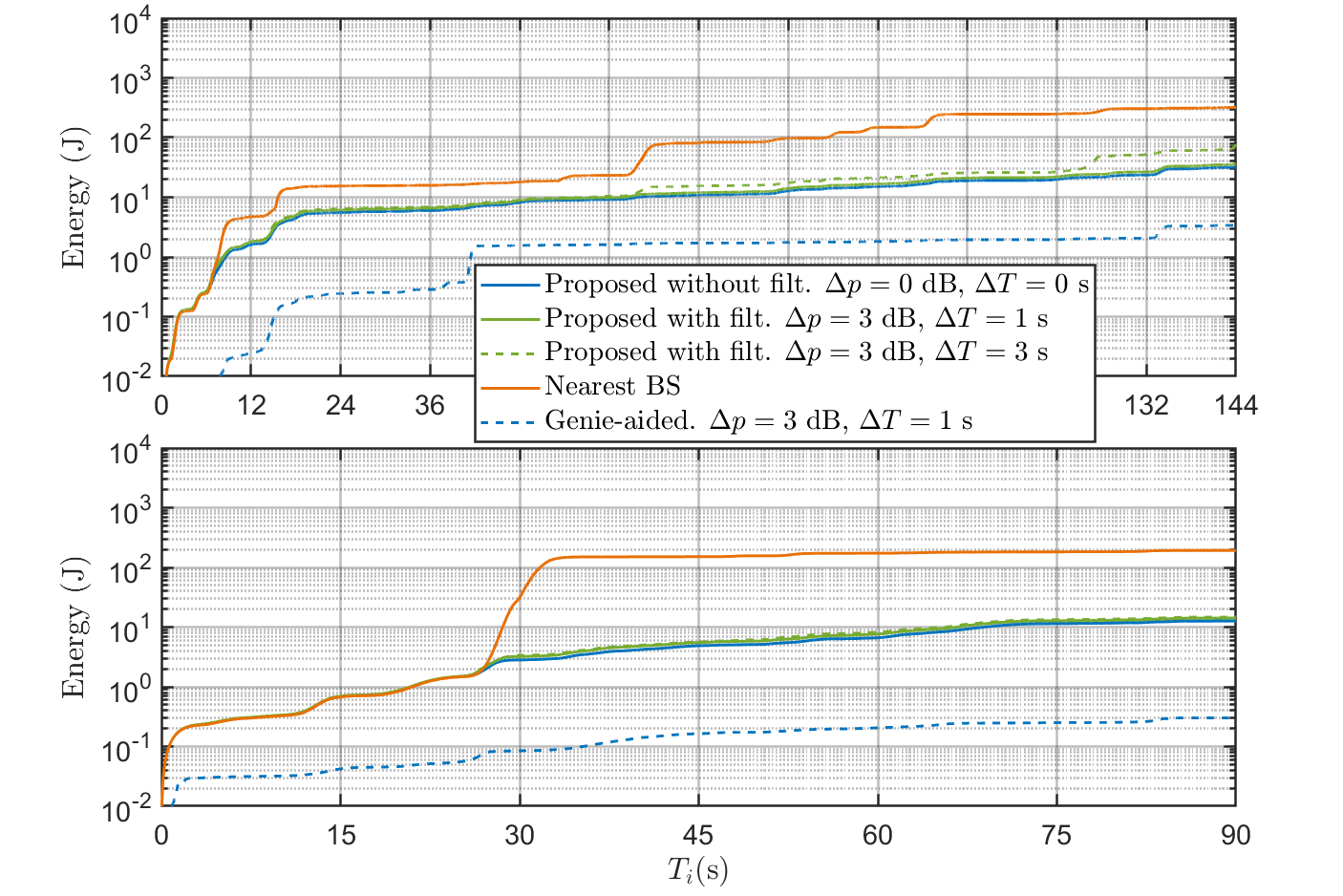}
    \caption{Energy consumption for $\zeta= 10^{-3}$. The figure on the top is for path 1 while the figure on the bottom is for path 2.}
    \label{Energy_cons_10dB}
\end{figure}

\begin{figure}[t!]
    \centering
    \includegraphics[width = \columnwidth]{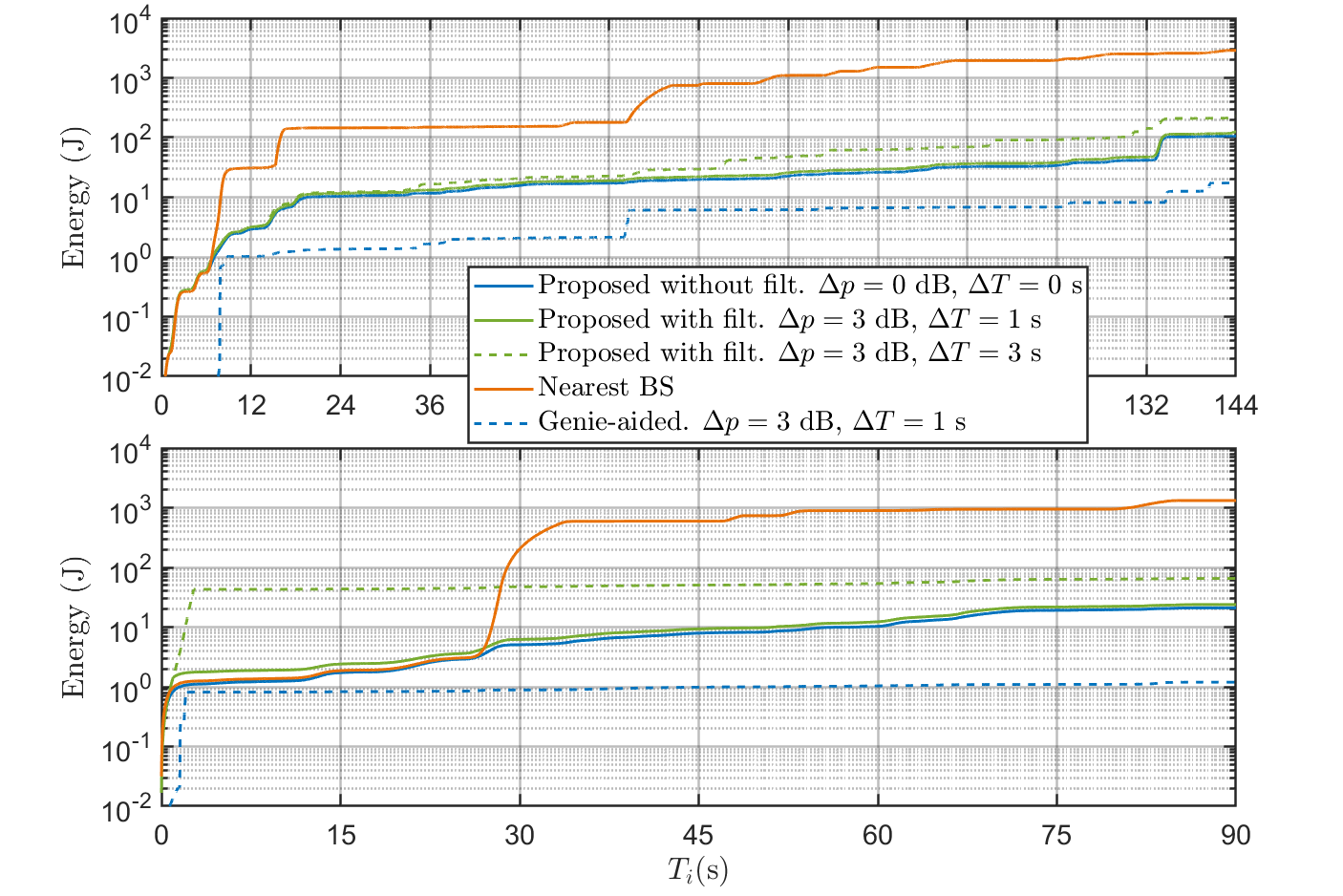}
    \caption{Energy consumption for $\zeta= 10^{-5}$. The figure on the top is for path 1 while the figure on the bottom is for path 2.}
    \label{Energy_cons_10dB_2}
\end{figure}

Fig. \ref{Energy_cons_10dB}  shows the energy consumed for transmission at the BSs as the UE moves along the paths for a target $\zeta= 10^{-3}$. Notice that the minimum energy consumption is achieved for a genie-aided approach with perfectly known SINR tail distribution. Getting close to the performance of this approach is extremely difficult since acquiring accurate knowledge of the channel distribution is challenging in practice. Note that its closer approach is our proposed scheme in \textbf{Algorithms} 1 and 2 with $\Delta p = 0$ dB and $\Delta T = 0$, which means that HOs are triggered only based on the UE position. Considering different values for $\Delta p$ and $\Delta T$ increases the energy cost but reduces the number of HOs for practical implementation. Notice that serving the UE by the nearest BS is more energy-demanding and does not represent the optimal strategy. Fig. \ref{Energy_cons_10dB_2} presents the same analysis but for a more strict outage probability target, \textit{e.g.,} $\zeta = 10^{-5}$.

\begin{figure}[t!]
    \centering
    \includegraphics[width = \columnwidth]{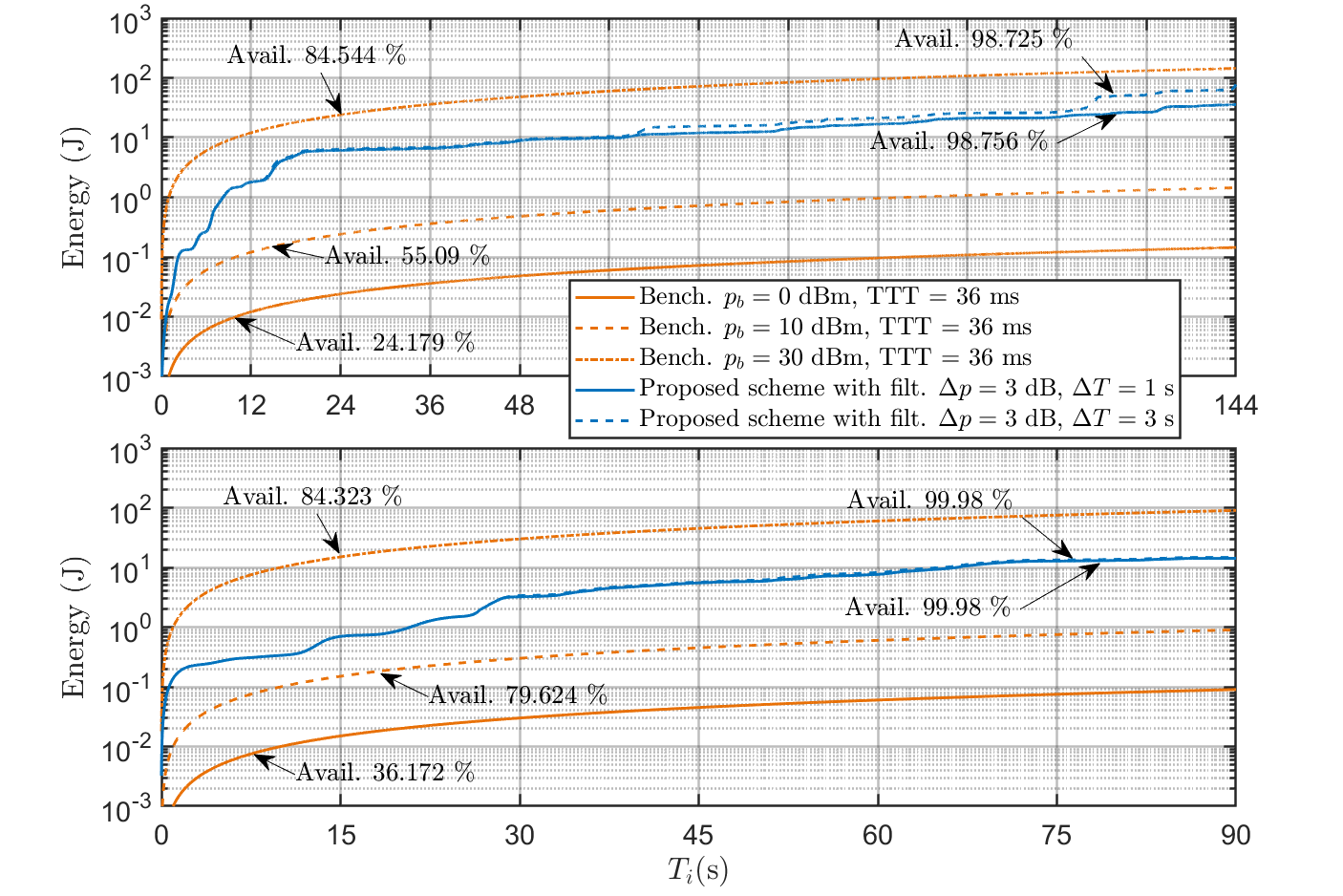}
    \caption{Comparison of the energy consumption and service availability with a conventional HO mechanism for $\zeta= 10^{-3}$.}
    \label{Energy_cons_10dB3}
\end{figure}

 Fig. \ref{Energy_cons_10dB3} depicts our proposed approach's energy consumption and service availability compared to a traditional HO mechanism across various $\Delta T$ values. Here, an HO decision is initiated when the SINR drops below the threshold $\gamma_0$ for a period exceeding the TTT. The HO is then executed towards the BS that offers the highest SINR at the time instant $T_i$. It is evident from the comparison that the benchmark model achieves lower energy consumption but at the expense of reduced service availability. This means that the outage probability frequently surpasses the target along significant trajectory segments. The benchmark doesn't outperform our approach in service availability even when operating at higher power levels. 
\section{Conclusions}\label{section6}
In this work, we proposed a new HO strategy and resource allocation exploiting radio maps and EVT  to support URC. The predictive capabilities of refined radio maps combined with the accuracy of EVT in modelling rare events allow accurate SINR characterisation across different spatial positions. Exploiting UE position information enabled the dynamic selection of the most suitable BS based on the information of the radio maps and time-power control strategies to avoid the ping-pong effect. The presented HO strategy was shown to achieve higher service availability in terms of outage probability and lower energy consumption when compared with classical HO mechanisms and the strategy that serves the UE from the nearest BS, respectively.

Notably, the approach presented in this work assumes perfect localisation of the UEs. However, this is a strong assumption, as random localisation error often exists in practice, even in high-accuracy systems. Nevertheless, the mathematical complexity of dealing with an additional RV can be significantly reduced if bound statistics are exploited. For instance, if the maximum localisation error distance is known, or an upper quantile, the allocated transmit power under localisation uncertainties would be the maximum transmit power within a circumference with the estimated coordinates as centre and radius equal to the maximum error/quantile. Finally, although the main scope of this work is oriented toward reliability, using a localisation-based approach for HO decisions instead of a continuous measure of QoS metrics would considerably reduce the HO time. This potentially reduces the hardware/signaling overhead in classic HO mechanisms, thus being relevant for URC.
\bibliographystyle{IEEEtran}
\bibliography{IEEEabrv,bibliography}

\end{document}